\documentstyle[12pt,emlines2,bezier,epsfig,aps,amsfonts]{revtex}
\tightenlines
\begin{document}
%
\title{Multifractality of Brownian Motion Near Absorbing Polymers}
\author{C.~von~Ferber$^{1,2}$ and Yu.~Holovatch$^3$}
\address{
$^1$ School of Physics and Astronomy, Tel Aviv University
IL-69978 Tel Aviv, Israel \\
$^2$ Institut f\"ur Theoretische Physik II, Heinrich-Heine-Universit\"at 
D\"usseldorf, D-40225 D\"usseldorf, Germany\\
$^3$Institute for Condensed Matter Physics,
Ukrainian Academy of Sciences,
UA-290011 Lviv, Ukraine
}
\maketitle
\date{\today}
\begin{abstract}
We characterize the multifractal behavior of Brownian motion
in the vicinity of an absorbing star Polymer. 
We map the problem to an
$O(M)$-symmetric $\phi^{4}$-field theory relating higher moments of the 
Laplacian field of Brownian motion to corresponding composite
operators. The
resulting spectra of scaling dimensions of these operators display the 
convexity properties which are necessarily found for multifractal scaling but
unusual for power of field operators in field theory.
Using a field-theoretic renormalization group approach we obtain the
multifractal spectrum for 
absorbtion at the core of a polymer star as an 
asymptotic series. We evaluate these series using  
resummation techniques. 
\end{abstract}
PACS {64.60.Ak, 61.41.+e, 64.60.Fr, 11.10.Gh}
\newpage
\section{Introduction}\label{I}
The concept of multifractality developed in the last decade has proven
to be a powerful tool for analyzing systems with complex statistics
which otherwise appear to be intractable \cite{Hentschel83,Halsey86}.
It has found direct application in a
wide range of fields including turbulence, chaotic attractors,
Laplacian growth phenomena
etc. 
\cite{mfract,growth}.
Let us give a simple example of a multifractal (MF) phenomenon. On a 
possibly fractal set $X \subset {\mathbb R}^d$ of total size $R$ 
a field $\varphi(r)$ is given at a microscopic 
scale $\ell$. Then normalized moments of this field may have power law
scaling behavior for $\ell/R\to 0$: 
\begin{equation}
\label{1.1}
<\varphi(r)^n>/<\varphi(r)>^n \sim (R/\ell)^{-\tau_n}.
\end{equation}
Nontrivial multifractal scaling is found if $\tau_n \neq 0$. When the moments 
are averages over the sites of $X$, $\varphi(r)$ defines a measure
on $X$ and rigorous arguments show that the 
$\tau_n$ are convex from  below as functions of $n$ 
\cite{Feller66,Duplantier91}. 

Here, we generalize an idea of Cates  and Witten \cite{Cates}
by deriving the MF spectrum in
the frames of a field theoretical formalism and make use of
renormalization group (RG) methods.  We  relate the MF spectrum  to
the spectrum of scaling dimensions of a family of composite operators
of Lagrangian $\phi^4$ field theory.
This gives an example of power of field
operators whose scaling dimensions show the appropriate
convexity for a MF spectrum \cite{Duplantier91,FerHol97,FerHol97b,FerHol97a}.

We thus address a special case of a  growth process controlled by a
Laplacian field.
The latter may describe a variety of phenomena depending on the
interpretation of the field.
For diffusion limited aggregation this
field is given by the concentration of diffusing particles, in
solidification processes it is given by the temperature field, in
dielectric breakdown it is the electric potential, in viscous fingers
formation it is the pressure \cite{growth}. In all
these processes the resulting structure appears to be of fractal
nature and is characterized by appropriate fractal
dimensions \cite{Mandelbrot83}. The growth and spatial
correlations of the structure are governed by spectra of
multifractal dimensions \cite{Hentschel83,Halsey86}. In general, the boundary
conditions determining the field will be given on the surface of the
growing aggregate itself. It is this dynamic coupling that produces the
rich structure of the phenomena and seems to make the general
dynamical problem intractable.

Here, we study a simpler case where the fractal structure is given and
we look for the distribution of a Laplacian field $\rho(r)$ 
and its higher
moments near the surface of the structure \cite{Cates}.
We will follow the diffusion
picture, considering the aggregate as an absorbing fractal,
``the absorber''. The field $\rho(r)$ gives the concentration of
diffusing particles and vanishes on the surface of the absorber.
More specifically, we consider the Laplacian field $\rho(r)$ in
the vicinity of an absorbing polymer, or near the core of a polymer star. 
In general, we assume
the ensemble of absorbers to be characterized by either random walk
(RW) or self-avoiding walk (SAW) statistics.
Multifractal scaling is found for the $n$-moments
$\langle \rho^n(r)\rangle$
of the field with respect to these ensembles.

This formulation of the problem allows us to use the polymer picture and
theory developed for polymer networks and
stars \cite{stars,Schaefer92} and
extended for copolymer stars \cite{FerHol97b,FerHol97a}. The theory
is mapped to a Lagrangian $\phi^4$ field theory with
several couplings \cite{Schaefer85,Schaefer90,Schaefer91}
and higher order composite operators \cite{Schaefer92,FerHol97b,FerHol97a} 
to describe star vertices.

Our article is organized in the following way. In the next section we 
present the path integral formulation of the Laplace equation and relate 
it to a polymer representation. The  field theoretical representation and 
renormalization of this polymer model is 
discussed in section \ref{III} where we discuss the renormalization group
(RG) flow and 
corresponding expressions for the exponents $\tau_n$. 
We calculate the multifractal spectrum to third order of
perturbation theory using two complementary approaches: the zero mass
renormalization with successive $\varepsilon$-expansion (see
e.g. \cite{Brezin76}) and the massive renormalization group approach at
fixed dimension \cite{Parisi80}. For some special cases
we reproduce previous results \cite{Cates}
that were obtained in lower order of perturbation theory. 
In section \ref{IV} we derive the multifractal spectra in terms of series 
expansions. These we present in both of our RG approaches.
The resulting series are asymptotic. In section \ref{V} we take this 
into account and obtain numerical values only by careful resummation.
In section \ref{VI} we discuss our results and conclude the present study.

\section{Path Integral Solution of the Laplace Equation and
polymer Absorber Model
}\label{II}
In this section we show how to describe the diffusion of
particles in the presence of a absorbing polymer using a ``polymer''
formalism that represents both the random walks of the diffusing
particles and the absorber itself in the same way \cite{Cates,Oshanin}. 
Let us formulate
this problem first in terms of diffusion of particles in time. The
probability to find a diffusing particle at point $r_1$
at time $t$ if it started at point $r_0$ at time $t=0$ can be
described by the following normalized path integral:
\begin{equation}\label{2.1}
G^{0}(r_0,r_1,t) = \langle \delta(r^{(1)}(0)-r_0)
\delta(r^{(1)}(t)-r_1)\rangle_{{\cal{H}}_0(r^{(1)},t)}.
\end{equation}
The angular brackets in Eq.\ (\ref{2.1}) stand for the following average:
\begin{equation}\label{2.2}
\langle \cdots \rangle_{{\cal{H}}_0(r^{(1)},t)} =
\frac{\int (\cdots) \exp (-  {\cal{H}}_0(r^{(1)},t)){\rm d} \{ r^{(1)} \}}
{ \int \exp (-  {\cal{H}}_0(r^{(1)},t))
{\rm d} \{ r^{(1)} \}},
\end{equation}
which is performed with the Hamiltonian:
\begin{equation}\label{2.3}
{\cal{H}}_0(r^{(1)},t)=\int_0^t\Big(\frac{{\rm d}r^{(1)}
(\tau)}{2 \, {\rm d} \tau}\Big )^2 {\rm d} \tau.
\end{equation}
The integration in Eq.\ (\ref{2.1}) is performed over all paths $r^{(1)}(\tau)$
with $0 \leq \tau \leq t$. Note that we have absorbed the diffusion
constant by a re-definition of time. The unit of the dimensionless Hamiltonian
${\cal H}_0$ is the product $k_BT$ of Boltzmann constant and temperature 
while that of time $t$ is the square microscopic 
length $\ell^2$. Spatial boundaries may be
included in Eq.\ (\ref{2.1}) by restricting the path integral to a subspace.
$G^{0}(r_0,r_1,t)$ obeys the
following differential equation:
\begin{equation}\label{2.4}
\Big( \Delta + \frac{d}{2t} - \frac{\partial}{\partial t} \Big )
G^{0}(r_0,r_1,t)  = 0.
\end{equation}
Here, $d$ is the dimension of space. In a given volume $V$ a
random walker will visit
any site after some finite time and we may assume a steady state limit
for $G^{0}(r_0,r_1,t)$ for $t=\infty$.
In this case $G^{0}(r_0,r_1,t)$ will become independent of
$r_0$ and its limit will define a field $\rho(r)$:
\begin{equation}\label{2.5}
\rho(r_1)=\lim_{t \rightarrow \infty }\frac {1}{V}
\int {\rm d} r_0 G^{0}(r_0,r_1,t).
\end{equation}
This field $\rho(r)$ then obeys the
Laplace equation:
\begin{equation}\label{2.6}
\Delta \rho (r)= 0.
\end{equation}

We introduce boundary conditions in such a way that the field
$\rho ( r)$ equals some constant $\rho_{\infty}$ at the extremal volume 
boundaries
(or at  $|r|=\infty$ ) and vanishes on the absorber. The absorber itself 
we describe
by a path $ r^{(2)}(s)$,
$0\leq s\leq S_2 $.
These boundary conditions are implemented by an
avoidance interaction $u_{12}$ punishing any coincidence of the path
$r^{(1)} $ of the RW and the path $ r^{(2)}$ of the absorber.
The correlation function of a random walk in the presence of an
absorbing path $r^{(2)}(s)$ with $0 \leq s \leq S_2$
may then be written as
\begin{eqnarray}
G(r_0,r_1,S_1)=\langle
\delta (r^{(1)}(0) - r_0)
\delta (r^{(1)}(S_1) - r_1)
\nonumber \\
\times \exp \Big \{
- \frac{u_{12}}{3!} \int_0^{S_1}{\rm d}s_1 \int_0^{S_2}{\rm d}s_2
 \delta (r^{(1)}(s_1) - r^{(2)}(s_2)
\Big \}
\rangle_{{\cal{H}}_0(r^{(1)},S_1)} ,
\label{2.10}
\end{eqnarray}
where we have adopted the notation $t=S_1$.

\begin{figure}
\makebox{\hspace*{30mm}\makebox{\input{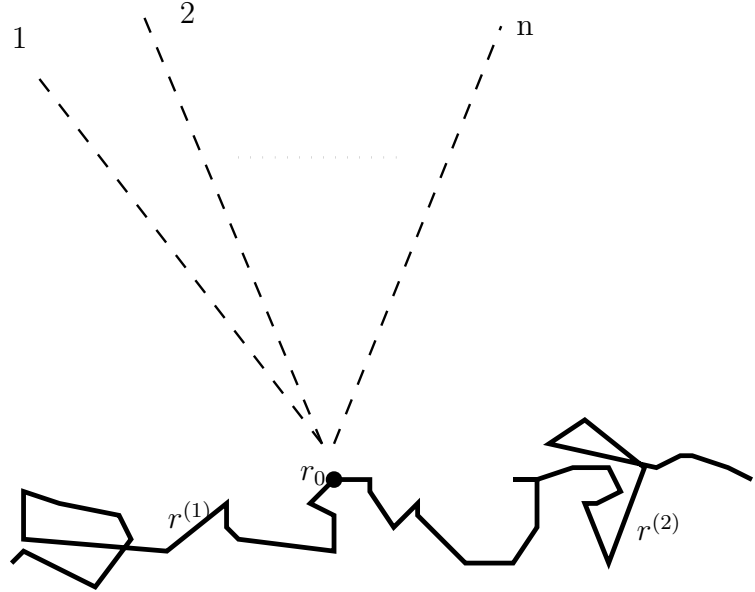}}}
\caption{\label{fig1.1}
   $n$ random walks that end at point $r_0$ on the absorbing polymer.}
\end{figure}

We are interested in ensemble averaged moments 
$\langle\rho^{n}(r_0+\xi)\rangle$
of the field in the vicinity of the absorber, i.e. with microscopic $\xi$.
For the RW ensemble the average is performed with respect to the
Hamiltonian ${\cal H}_0(r^{(2)},S_2)$, for the SAW ensemble an additional
interaction has to be included.
The moments we calculate as an ensemble average over all 
configurations of the absorbing polymer choosing the site $r_0$
on the middle of the polymer \cite{note1}. Formally we write these moments for
$\xi\to 0$ as
\begin{equation}
\lim_{|\xi|\to 0}\langle\rho^{n}(r_0+\xi)\rangle = \lim_{S_{a>m}\to\infty}
\frac{1}{{\cal Z}^0_{*m0}} \int \prod_{a=1}^{m+n}{\rm d}r_a\, 
G_{mn}^*(r_0,r_1, \ldots,r_{m+n},S_1,\ldots,S_{m+n}).
\label{2.10a}
\end{equation}
The normalization ${\cal Z}^0_{*m0}$ takes care of the configurations of the 
absorber,
as explained in the next section. The correlation function $G_{mn}^*$
is defined as 
\begin{eqnarray}
G_{mn}^*(r_0,r_1, \ldots,r_{m+n},S_1,\ldots,S_{m+n})
=\langle
 \prod_{a=1}^{m+n}
\delta (r^{(a)}(0) - r_0)
\delta (r^{(a)}(S_a) - r_a)
\nonumber \\
\times \exp \left\{-\sum_{a,b=1}^{m+n}
\frac{\bar{u}_{ab}}{3!} \int_0^{S_a}{\rm d}s_a \int_0^{S_b}{\rm d}s_b
 \delta (r^{(a)}(s_a) - r^{(b)}(s_b)
\right\}
\rangle_{\sum_a{\cal{H}}_0(r^{(a)},S_a)} .
\label{2.10b}
\end{eqnarray}
Here, the absorbing walk is represented by $m=2$ paths $r^{(1)}$,
$r^{(2)}$, while the remaining $n$ paths represent $n$ random walks,
as it is shown in the figure \ref{fig1.1}.
The interaction matrix $\bar{u}_{ab}$ is in this case given by 
$\bar{u}_{ab}=\{0 \; \mbox{if} \; a,b \leq m \; \mbox{or}\; a,b>m; 
\; u_{12}\; \mbox{else}\}$.

The limits in Eq.\ (\ref{2.10a}) look rather ill defined at first sight, 
and indeed they should not be taken naively. Also the evaluation of the 
functional integral (\ref{2.10b}) is not defined in this bare form. 
Luckily we have at hand the polymer field theory which has dealt 
with the problems of evaluating these formal expressions \cite{books}.
We will show below how the theory is mapped to a renormalizable $O(M)$
symmetric field theory in terms of which the limits and a perturbative 
expansion of (\ref{2.10b}) make sense. For instance the limit $|\xi|\to 0$
may be interpreted as a short distance limit defining a composite 
operator, while the limit $S_{a>m} \to \infty$, with $S_{b \leq m}$
stay finite corresponds to a short chain limit derived in 
\cite{Ferber97}. In the frames of the polymer picture we may interpret 
$G_{mn}^*(r_0,r_1, \ldots,r_{m+n},S_1,\ldots,S_{m+n})$ as the correlation 
function of $m+n$ interacting walks all starting at point $r_0$ with end 
points at $r_1, \ldots,r_{m+n}$. These describe what is called a polymer 
star. The normalized partition function of such a star of $m+n$  polymer 
chains with chain lengths parametrized by $S_a$ may be written as 
\cite{stars,Schaefer92}:
\begin{equation}
\label{2.11}
{\cal Z}_{*mn}\{S_a\}= \frac{1}{{\cal N}_{mn}}
\int \prod_{a=1}^{m+n}{\rm d}r_a\, 
G_{mn}^*(r_0,r_1, \ldots,r_{m+n},S_1,\ldots,S_{m+n}).
\end{equation}
The normalization ${\cal N}_{mn}$ is chosen such that 
${\cal Z}_{*mn}\{S_a\}|_{\bar{u}_{ab}=0}=1$ for vanishing interactions and the
point $r_0$ is arbitrary.  We have studied this problem of polymer stars 
with general interaction matrix $\bar{u}_{ab}$ in 
\cite{FerHol97b,FerHol97a}. Here we
will choose
\begin{equation}\label{2.12}
\bar{u}_{ab} = \left \{ \begin{array}{cc}
u_{11} & \mbox{if $a,b \leq m$}
\\
u_{22} & \mbox{if $a,b > m$}
\\
u_{12} & \mbox{else}
\end{array} \right.
\end{equation}
This allows us to think of the absorbing paths $r^{(1)},\ldots,r^{m}$
as being either of a RW ($u_{11}=0$) or of a SAW ($u_{11}\neq 0$) ensemble.
With $m>2$ we also include in our study the moments of the diffusion field 
near to the core of a polymer star. We included $u_{22}$ only to ease 
notation, in the present context $u_{22}=0$. 

\section{Field Theory and Renormalization
}\label{III}
As  is well known, the polymer model may be mapped to the limit
$M=0$ of  $O(M)$-symmetrical Lagrangian field theory \cite{deGennes72}.
We adopt the formalism developed for multicomponent
polymer solutions which allows us to describe both polymers and interacting 
random walks \cite{Schaefer91}. Its field theory  is given
by the following Lagrangian:
\begin{eqnarray}\label{3.1}
\lefteqn{ {\cal L}\{\phi_a,\mu_a\} = \frac{1}{2}
\sum_{a=1}^{m+n}\int{\rm d}^d r \left(\mu_a\phi_a^2 + (\nabla \phi_a(r)
)^2 \right)}&&\nonumber\\ &&+ \frac{1}{4!} \sum_{a,b=1}^{m+n}
\bar{u}_{a,b} \int {\rm d}^d r \phi_a^2(r)\phi_b^2(r) .
\end{eqnarray}
The $\phi^2$ terms should be read as scalar products of 
fields $\phi_a$ with $M$ component
\begin{equation}\label{3.2} 
\phi_a^2 = \sum_{\alpha = 1}^{M}(
  \phi_a^{\alpha} )^2 .  
\end{equation}
The parameter
$\mu_a$ is a chemical potential conjugated to the chain length variables
$S_a$ in Eq.\ (\ref{2.10}). Correlation functions in this theory are
defined by averaging with the Lagrangian ${\cal L}$:
\begin{equation} \label{3.3}
\langle (\cdots) \rangle_{{\cal L}}
=   \int{\cal D}[\phi_a(r)] (\ldots)
   \exp[-{\cal L}\{\phi_a,\mu_a\}] \, |_{M=0}.
\end{equation}
Here, functional integration $\int{\cal D}[\phi_a(r)]$ is defined in
such a way that normalization is already included:
$\langle 1 \rangle |_{{\cal L}} = 1 $ if all $\bar{u}_{a,b}\equiv0$.
The limit $M=0$ in Eq.\ (\ref{3.3}) can be understood as a selection rule
for the diagrams which contribute to the perturbation theory expansion
and can be easily checked in this context to correspond to interacting 
polymers in the following way:
The partition function ${\cal Z}_{*mn}$ defined  in Eq.\ (\ref{2.10}) is 
mapped 
to the field theoretical correlation function $\tilde{\cal Z}_{*mn}$ via 
Laplace
transforms in the chain length variables $S_a$ to conjugate chemical potentials
(``mass variables'') $\mu_a$:
\begin{equation} \label{3.4}
\tilde{\cal Z}_{*mn}\{\mu_a\} = \int _0^{\infty}
\prod_b {\rm d}S_b e^{-\mu_b S_b} {\cal Z}_{*mn}\{ S_a \} .
\end{equation}
In terms of the above defined Lagrangian field theory  $\tilde{\cal Z}_{*mn}$ 
is given by
\begin{equation} \label{3.5}
\tilde{\cal Z}_{*mn}\{\mu_a\} = \langle
\int  \prod_{a=1}^{m+n} {\rm d} r_a \phi_a(r_0) \phi_a(r_a)
\rangle |_{{\cal L}}.
\end{equation}

Our interest is in the scaling properties of these
functions.
Note that by (\ref{3.5}) these are governed by the spectrum of scaling
dimensions of the composite operators $\prod_{a=1}^{m+n} \phi_a$.
To extract these dimensions we use RG
methods \cite{Bogoliubov59,Zinn89}. Here, we apply  the results of
our previous approaches  to the problem of co-polymer stars
\cite{FerHol97b,FerHol97a}: massless
renormalization group scheme with successive $\varepsilon$-expansion
(see e.g.\cite{Brezin76}) and massive renormalization group approach
at fixed dimension \cite{Parisi80} compiled in a
pseudo-$\varepsilon$ expansion \cite{pseps}.
 On the basis of
correlation functions it is standard to define vertex functions
$\Gamma^{(4)}_{\bar{u}_{ab}}$ corresponding to the couplings $\bar{u}_{ab}$ 
as well as vertex functions $\Gamma^{*mn}_{\Pi \phi_a}$ with insertion of
composite operators $\prod_a \phi_a$. Explicit
expressions may be found in \cite{FerHol97a}. We define
renormalization and introduce renormalized couplings $\bar{g}_{ab}$ by:
\begin{equation} \label{3.6}
\bar{u}_{ab} = \mu^{\varepsilon} Z_{\phi_a}Z_{\phi_b}Z_{ab} \bar{g}_{ab} .
\end{equation}
The renormalizing $Z$-factors are power series in the renormalized couplings
$\bar{g}_{ab}$ subject to the following renormalization conditions:
\begin{eqnarray}
Z_{\phi_a}(\bar{g}_{aa}) \frac{\partial}{\partial k^2}
\Gamma_{aa}^{(2)}(\bar{u}_{aa}(\bar{g}_{aa})) = 1,             \label{3.7}\\
Z_{ab}(\{\bar{g}_{ab}\})
\Gamma_{aabb}^{(4)}(\bar{u}_{ab}(\{\bar{g}_{ab}\})) = \mu^\varepsilon 
\bar{g}_{ab}.
\label{3.8}
\end{eqnarray}
The scale parameter $\mu$
is equal to the mass at which the massive
scheme is evaluated or it gives the scale of the external momenta in the
massless scheme.

In order to renormalize the star vertex functions we introduce
renormalization factors  $Z^{*mn}_{\Pi \phi_a}$ by:
\begin{equation} \label{3.9}
(\prod_{a=1}^{k} Z_{\phi_a}^{1/2}) Z^{*mn}_{\Pi \phi_a}
\Gamma^{*mn}_{\Pi \phi_a}(\bar{u}_{ab}(\{\bar{g}_{ab}\})) = \mu^{\delta_{\Pi 
\phi_a}},
\end{equation}
where $\delta_{\Pi \phi_a}$ is the engineering dimension of the
composite operator
\begin{equation} \label{3.10}
\delta_{\Pi \phi_a}\ = \ (m+n)(\frac {\varepsilon}{2} -1)
+ \ 4 - \ \varepsilon.
\end{equation}
The renormalized couplings $\bar{g}_{ab}$ and renormalizing $Z$-factors depend 
on the scale parameter $\mu$. This is expressed by the following RG flow 
equations:
\begin{eqnarray}
\mu \frac{\rm d}{{\rm d}\mu} \bar{g}_{ab} &=& \bar{\beta}_{ab}
(\{\bar{g}_{ab}\}) ,
\label{3.11} \\
\mu \frac{\rm d}{{\rm d}\mu} \ln Z^{*mn}_{\Pi \phi_a}(\{\bar{g}_{ab}\}) &=&
\eta_{\Pi \phi_a}(\{\bar{g}_{ab}\}) .
\label{3.12}
\end{eqnarray}
Our original problem is described by two sets of walks of different 
species. In this case only the three different couplings 
$u_{11}$, $u_{12}$ and $u_{22}$ in Eq.\ (\ref{2.12}) remain.
We will refer to their renormalized counterparts 
as $g_{11}$, $g_{22}$, $g_{12}=g_{21}$.
The corresponding functions $\beta_{11}$, $\beta_{22}$, $\beta_{12}$ define
the flow in the space of couplings. This RG flow was
discussed in \cite{Schaefer90,Schaefer91}. Its fixed points are
determined by the set of equations:
\begin{eqnarray} 
\beta_{11}(g_{11}^*) &=& 0, \nonumber \\
\beta_{22}(g_{22}^*) &=& 0, \nonumber \\
\label{3.13}
\beta_{12}(g_{11}^*,g_{22}^*,g_{12}^*)&=&0.
\end{eqnarray}

In the space of the three couplings one finds \cite{Schaefer91} 8
fixed points corresponding to the absence or presence of inter- and intra-
species interaction. 
The equations for the fixed
points of the $\beta$-functions were found to
have the following nontrivial solutions:
$\beta_{{aa}}(g^*_{\rm S})= 0$
and for $a\neq b $:
$\beta_{{12}}(0          ,0          , g^*_{\rm G}) = 0$,
$\beta_{{12}}(g^*_{\rm S},0          , g^*_{\rm U}) = 0$,
$\beta_{{12}}(0          ,g^*_{\rm S}, g^*_{\rm U}) = 0$,
$\beta_{{12}}(g^*_{\rm S},g^*_{\rm S}, g^*_{\rm S}) = 0$,
 corresponding to all combinations of
interacting and non-in\-ter\-ac\-ting chains. 

The phenomenon we address in this article corresponds to the case of
a non-vanishing interaction between the two species of walks,
while one set has no self-interaction.
 Thus we consider the
two fixed points which we call {\it G} ($g_{11}=g_{22}=0, g_{12}=
g^*_{G}$)
and {\it U} ($g_{11}=g^*, g_{22}=0, g_{12}=
g^*_{U}$). The first ({\it G}) corresponds to a set of random walks
interacting with another set of random walks of a second species and thus 
describes
absorption on random walk absorbers, the second ({\it U}) corresponds to a
set of random walks interacting with a set of self-avoiding walks
and thus describes absorption on SAW (polymer) absorbers.

Having $m$ walks of the first species and $n$ walks of second
species we define the following exponents in the fixed points
{\it G,U}:
\begin{eqnarray}
\label{3.14}
\eta^{G}_{mn}&=& \eta_{\Pi \phi_a}(g_{11}=g_{22}=0, g_{12}=
g^*_{G}),
\\
\label{3.15}
\eta^{U}_{mn}&=& \eta_{\Pi \phi_a}(g_{11}=g^*, g_{22}=0, g_{12}=
g^*_{U}),
\end{eqnarray}
which govern the scaling properties of the partition sum (\ref{2.11}).

The scaling may be formulated in terms of the size $R$ of the absorbing walks
while the RWs of the diffusing particles are taken infinitely long. This 
corresponds
to a short chain expansion \cite{Ferber97}.
We have to normalize the
partition function by the number of configurations of the absorber
given by  ${\cal Z}_{*m0}$ and by the $n$th power of the first moment (see Eq.
(\ref{1.1})).
For large $R$ on the microscopic scale $\ell$ 
the moments of $\rho(r_0)$ at 
point $r_0$ in the vicinity of the core of the star scale like 
\begin{equation}
\label{3.16}
\frac{<\rho(r_0)^n>}{<\rho(r_0)>^n}=
{\cal Z}_{*mn}/{\cal Z}_{*m0} 
\Big( {\cal Z}_{*m1}/{\cal Z}_{*m0} \Big)^{-n}\sim \Big(
\frac{R}{\ell}\Big)^{-\tau_{mn}},
\end{equation}
where $R=S_{a\leq m}^{\nu}$ and the exponents $\tau_{mn}$ are given as
\begin{equation}
\label{3.17}
\tau_{mn}=-\eta_{mn}+n\eta_{m1}-(n-1)\eta_{m0}.
\end{equation}
Here, $\nu$ is
the correlation length critical exponent of the walks:
$\nu=1/2$ for random walks and $\nu \simeq 0.588$ for self-avoiding
walks at $d=3$. For the fixed point {\it G} we have $\eta_{m0}^G=0$ and
$\nu=1/2$ for all walks. The scaling near a RW star is then described by 
inserting 
the value of $\eta^G$ while the scaling near a SAW star is obtained by 
inserting 
$\eta^U$. In previous work \cite{FerHol97b,FerHol97a}, we obtained
the exponents $\eta^{G}_{mn}$,
$\eta^{U}_{mn}$ in third order of perturbation theory.

\section{Multifractal Spectrum}\label{IV}
A widely used characterization for the MF spectrum is the so called spectral 
function
$f_m(\alpha)$ \cite{Halsey86}. To obtain this function for the absorption 
process
on the center of a star with $m$ legs we analytically continue the set of 
exponents 
$\tau_{mn}$ in the variable $n$ and calculate the following Legendre transform
\begin{equation}
\label{4.1}
f_m(\alpha_{mn})= -\tau_{mn} +n\alpha_{mn} + D_m
\mbox{\hspace{2em} with \hspace{2em}}
\alpha_{mn}=\frac{{\rm d}\tau_{mn}}{{\rm d}n}+D_m.
\end{equation}
Following the standard definition we have included into Eq.\ (\ref{4.1}) 
the fractal 
dimension  $D_m$ of the absorber. In particular, this gives the maximal value 
of the
spectral function $f_m(\alpha_{mn})$ to be equal to the dimension $D_m$. 
However 
the 
last is ill defined in the case of absorption near 
the 
core of a polymer star. That is why in the subsequent analysis we
will neglect $D_m$ in the expressions (\ref{4.1}).
Note as well that we have defined the exponents $\tau_{mn}$ based on ensemble 
averages for the moments $<\rho(r)^m>$. This definition deviates from 
the standard approach where an average over the support of $\rho(r)$
is assumed. One has to expect some deviations from the behavior of the standard
$f(\alpha)$ which we will discuss below \cite{Cates}.  
To obtain the expressions for the spectral function we use the perturbation 
expansions for the $\eta$ exponents given to third order both in massless and 
massive renormalization \cite{FerHol97b,FerHol97a}. These exponents are 
available 
both in terms of $\varepsilon$-expansion and pseudo-$\varepsilon$-expansion
series. 
The first
corresponds to collecting perturbation theory terms of the same power
of $\varepsilon=4-d$. In the pseudo-$\varepsilon$ expansion
series \cite{pseps} each power of the
pseudo-$\varepsilon$ parameter $\tau$
collects the contributions from the dimension-dependent loop integrals
of the same order.
In the final results the limit $\tau=1$ is taken.
Starting from the relations for $\tau_{mn}$ (\ref{3.17}) and the spectral 
function (\ref{4.1}) some algebra results in the corresponding expansions 
for the MF spectra for absorption on stars of random walks (RW) and self 
avoiding walks (SAW):
\begin{eqnarray}
\label{4.2}
\alpha_{mn}^{RW}(\varepsilon)&=&
-m\Big (2n-1\Big ){\varepsilon}^{2}
/8
+m\Big (4mn+6n\zeta (3)-12n+3{n}^{2}
\\&&\nonumber
+5-2m-3\zeta (3)
\Big ){\varepsilon}^{3}
/16,
\\ 
\label{4.3}
f_m^{RW}(\varepsilon)&=&
-m{n}^{2}{\varepsilon}^{2}
/8
+m{n}^{2}\Big (-6+2n+2m+3\zeta (3)\Big ){\varepsilon}^{3}
/16,
\\ \label{4.4}
\alpha_{mn}^{SAW}(\varepsilon)&=&
-9m\Big (2n-1\Big ){\varepsilon}^{2}
/128
+3m\Big (168mn+54{n}^{2}+157
\\&&\nonumber
+180n\zeta (3)
-350n-84m-90\zeta (3)\Big ){\varepsilon}^{3}
/2048,
\\ \label{4.5}
f_m^{SAW}(\varepsilon)&=&
-9m{n}^{2}{\varepsilon}^{2}
/128
+3m{n}^{2}\Big (-175+36n+84m+90\zeta (3)\Big ){\varepsilon}^{3}
/2048,
\\ \label{4.6}
\alpha^{RW}_{mn}(\tau)&=&
-m\varepsilon\Big (1+4ni_{{1}}-2i_{{1}}-2n\Big ){\tau}^{2}
/4
+ \alpha^{RW}_{3loop}{\tau}^{3},
\\ \label{4.7}
f_m^{RW}(\tau)&=&
-\varepsilon m{n}^{2}\Big (-1+2i_{{1}}\Big ){\tau}^{2}/4
+ f^{RW}_{3loop}{\tau}^{3},
\\ \label{4.8}
\alpha_{mn}^{SAW}(\tau)&=&
-9m\varepsilon\Big (1-2n+4ni_{{1}}-2i_{{1}}\Big ){\tau}^{2}
/64
+ \alpha^{SAW}_{3loop}{\tau}^{3},
\\ \label{4.9}
f_m^{SAW}(\tau)&=&
-9\varepsilon m{n}^{2}\Big (-1+2i_{{1}}\Big ){\tau}^{2}/64
+ f^{SAW}_{3loop}{\tau}^{3}.
\end{eqnarray}
Here $\zeta(3) \simeq 1.202$ is the Riemann zeta function,
$i_1$, $i_2$ are the two-loop  integrals depending on the space dimension 
$d$: at $d=3$ $i_1=2/3$, $i_2=-2/27$. The explicit form of the three-loop
contributions in Eqs.\ (\ref{4.6})--(\ref{4.9}) is given in the 
appendix \ref{appA}.

\section{Resummation and Results
}\label{V}

As is well known, the series of type
(\ref{4.2}) - (\ref{4.9}), as they occur in field theory appear
to be of asymptotic nature with zero radius of convergence. However,
knowing the asymptotic behavior of the series as
derived from the RG theory we may evaluate these
asymptotic series by resummation (see e.g. \cite{LeGuillou80}). 
To this end several procedures are available differing in the amount 
of information that is used to control the convergence. We extract this
additional information  for the case of
our Lagrangian  (\ref{3.1}) from 
\cite{Schaefer91,ass}. 
We expect the
following behavior of the $k$th order perturbation theory term $A_k$
for any of the above quantities:  
\begin{equation} \label{5.0} 
A_k\,\sim \,k!\,k^b\,(-a)^k .
\end{equation} 
The constant $a$ for the
$\varepsilon$-expansion of Lagrangian $\phi^4$ field theory with one
coupling was derived in  \cite{ass}:  $a=3/8$. For the
unsymmetric fixed point $U$, where two different couplings are present
the value $a=27/64$ has been proposed \cite{Schaefer91}. We assume here that 
the
same properties also hold for the pseudo-$\varepsilon$ expansion in
terms of $\tau$. With the above information at hand one can make use
of the Borel summation technique improved by the conformal mapping
procedure which has served a powerful tool in field theory calculations
(see \cite{LeGuillou80} for example).
We present some details of the resummation procedure in appendix \ref{appB}.

As stated before, we have chosen a definition for the generalized spectral
functions $f_m(\alpha)$ neglecting the fractal dimension $D_m$ of the absorber.
In this way we ease the comparison of the 
spectral functions $f_m(\alpha)$ for different values of $m$, the number
of legs of the absorbing star. We have shifted all maxima of $f_m(\alpha)$
to the point known for $m=2$ with $D_{m=2}=2$ in the RW case and $D_{m=2}=1.71$
in the SAW case.

\begin{figure}
\makebox{\hspace*{-20mm}\makebox{
\setlength{\unitlength}{0.1bp}
\special{!
/gnudict 40 dict def
gnudict begin
/Color false def
/Solid false def
/gnulinewidth 5.000 def
/vshift -33 def
/dl {10 mul} def
/hpt 31.5 def
/vpt 31.5 def
/M {moveto} bind def
/L {lineto} bind def
/R {rmoveto} bind def
/V {rlineto} bind def
/vpt2 vpt 2 mul def
/hpt2 hpt 2 mul def
/Lshow { currentpoint stroke M
  0 vshift R show } def
/Rshow { currentpoint stroke M
  dup stringwidth pop neg vshift R show } def
/Cshow { currentpoint stroke M
  dup stringwidth pop -2 div vshift R show } def
/DL { Color {setrgbcolor Solid {pop []} if 0 setdash }
 {pop pop pop Solid {pop []} if 0 setdash} ifelse } def
/BL { stroke gnulinewidth 2 mul setlinewidth } def
/AL { stroke gnulinewidth 2 div setlinewidth } def
/PL { stroke gnulinewidth setlinewidth } def
/LTb { BL [] 0 0 0 DL } def
/LTa { AL [1 dl 2 dl] 0 setdash 0 0 0 setrgbcolor } def
/LT0 { PL [] 0 1 0 DL } def
/LT1 { PL [4 dl 2 dl] 0 0 1 DL } def
/LT2 { PL [2 dl 3 dl] 1 0 0 DL } def
/LT3 { PL [1 dl 1.5 dl] 1 0 1 DL } def
/LT4 { PL [5 dl 2 dl 1 dl 2 dl] 0 1 1 DL } def
/LT5 { PL [4 dl 3 dl 1 dl 3 dl] 1 1 0 DL } def
/LT6 { PL [2 dl 2 dl 2 dl 4 dl] 0 0 0 DL } def
/LT7 { PL [2 dl 2 dl 2 dl 2 dl 2 dl 4 dl] 1 0.3 0 DL } def
/LT8 { PL [2 dl 2 dl 2 dl 2 dl 2 dl 2 dl 2 dl 4 dl] 0.5 0.5 0.5 DL } def
/P { stroke [] 0 setdash
  currentlinewidth 2 div sub M
  0 currentlinewidth V stroke } def
/D { stroke [] 0 setdash 2 copy vpt add M
  hpt neg vpt neg V hpt vpt neg V
  hpt vpt V hpt neg vpt V closepath stroke
  P } def
/A { stroke [] 0 setdash vpt sub M 0 vpt2 V
  currentpoint stroke M
  hpt neg vpt neg R hpt2 0 V stroke
  } def
/B { stroke [] 0 setdash 2 copy exch hpt sub exch vpt add M
  0 vpt2 neg V hpt2 0 V 0 vpt2 V
  hpt2 neg 0 V closepath stroke
  P } def
/C { stroke [] 0 setdash exch hpt sub exch vpt add M
  hpt2 vpt2 neg V currentpoint stroke M
  hpt2 neg 0 R hpt2 vpt2 V stroke } def
/T { stroke [] 0 setdash 2 copy vpt 1.12 mul add M
  hpt neg vpt -1.62 mul V
  hpt 2 mul 0 V
  hpt neg vpt 1.62 mul V closepath stroke
  P  } def
/S { 2 copy A C} def
end
}
\begin{picture}(5039,3023)(0,0)
\special{"
gnudict begin
gsave
50 50 translate
0.100 0.100 scale
0 setgray
/Helvetica findfont 100 scalefont setfont
newpath
-500.000000 -500.000000 translate
LTa
600 251 M
4256 0 V
600 251 M
0 2721 V
LTb
600 251 M
63 0 V
4193 0 R
-63 0 V
600 523 M
63 0 V
4193 0 R
-63 0 V
600 795 M
63 0 V
4193 0 R
-63 0 V
600 1067 M
63 0 V
4193 0 R
-63 0 V
600 1339 M
63 0 V
4193 0 R
-63 0 V
600 1612 M
63 0 V
4193 0 R
-63 0 V
600 1884 M
63 0 V
4193 0 R
-63 0 V
600 2156 M
63 0 V
4193 0 R
-63 0 V
600 2428 M
63 0 V
4193 0 R
-63 0 V
600 2700 M
63 0 V
4193 0 R
-63 0 V
600 2972 M
63 0 V
4193 0 R
-63 0 V
600 251 M
0 63 V
0 2658 R
0 -63 V
1132 251 M
0 63 V
0 2658 R
0 -63 V
1664 251 M
0 63 V
0 2658 R
0 -63 V
2196 251 M
0 63 V
0 2658 R
0 -63 V
2728 251 M
0 63 V
0 2658 R
0 -63 V
3260 251 M
0 63 V
0 2658 R
0 -63 V
3792 251 M
0 63 V
0 2658 R
0 -63 V
4324 251 M
0 63 V
0 2658 R
0 -63 V
4856 251 M
0 63 V
0 2658 R
0 -63 V
600 251 M
4256 0 V
0 2721 V
-4256 0 V
600 251 L
LT0
4553 2809 M
180 0 V
3721 251 M
-69 696 V
-103 657 V
-92 407 V
-80 272 V
-73 191 V
-65 138 V
-58 102 V
-54 76 V
-48 57 V
-45 43 V
-41 31 V
-38 22 V
-35 15 V
-32 9 V
-30 5 V
-28 0 V
-27 -4 V
-24 -6 V
-23 -9 V
-21 -12 V
-27 -13 V
-15 -15 V
-18 -16 V
-18 -18 V
-16 -18 V
-16 -20 V
-14 -21 V
-15 -22 V
-13 -23 V
-13 -23 V
-12 -24 V
-12 -24 V
-11 -25 V
-11 -26 V
-10 -26 V
-10 -26 V
-10 -27 V
-9 -27 V
-9 -27 V
-8 -28 V
-9 -28 V
-8 -28 V
-7 -29 V
-7 -29 V
-8 -29 V
-6 -29 V
-7 -29 V
-7 -30 V
-6 -30 V
-6 -30 V
-6 -30 V
-5 -30 V
-6 -30 V
-5 -31 V
-5 -30 V
-5 -31 V
-5 -31 V
-5 -31 V
-5 -31 V
-4 -31 V
-4 -31 V
-5 -32 V
-4 -31 V
-4 -32 V
-4 -31 V
-4 -32 V
-3 -31 V
-4 -32 V
-4 -32 V
-3 -32 V
-4 -32 V
-3 -32 V
-3 -32 V
-3 -32 V
-3 -32 V
-3 -32 V
-3 -33 V
-3 -32 V
-3 -32 V
-3 -33 V
-3 -32 V
-2 -32 V
-3 -33 V
-2 -32 V
-3 -33 V
-2 -33 V
-3 -32 V
-2 -33 V
-2 -32 V
-3 -33 V
-2 -33 V
-2 -33 V
-2 -32 V
-2 -33 V
-2 -33 V
-2 -33 V
-2 -33 V
-2 -33 V
-2 -33 V
-2 -33 V
-2 -33 V
-2 -33 V
-2 -33 V
-1 -33 V
-2 -33 V
-2 -33 V
-1 -33 V
-2 -33 V
-2 -33 V
-1 -33 V
-1 -22 V
LT1
4553 2709 M
180 0 V
3964 251 M
-84 502 V
-104 496 V
-94 373 V
-86 287 V
-79 226 V
-72 180 V
-67 144 V
-63 116 V
-57 94 V
-54 75 V
-51 61 V
-47 49 V
-44 38 V
-41 29 V
-39 22 V
-37 15 V
-35 9 V
-33 4 V
-31 0 V
-30 -3 V
-28 -8 V
-26 -10 V
-25 -13 V
-29 -16 V
-21 -18 V
-22 -19 V
-21 -22 V
-20 -23 V
-20 -25 V
-18 -27 V
-18 -27 V
-17 -29 V
-17 -30 V
-16 -31 V
-15 -32 V
-15 -32 V
-15 -34 V
-13 -34 V
-14 -35 V
-13 -36 V
-12 -36 V
-12 -37 V
-12 -37 V
-11 -38 V
-11 -39 V
-11 -38 V
-10 -40 V
-10 -39 V
-10 -40 V
-9 -41 V
-9 -41 V
-9 -41 V
-9 -41 V
-8 -42 V
-8 -42 V
-8 -42 V
-8 -42 V
-8 -43 V
-7 -43 V
-7 -43 V
-7 -43 V
-7 -44 V
-7 -44 V
-6 -44 V
-7 -44 V
-6 -44 V
-6 -45 V
-6 -44 V
-5 -45 V
-6 -45 V
-6 -45 V
-5 -45 V
-5 -45 V
-6 -46 V
-5 -45 V
-5 -46 V
-4 -46 V
-5 -45 V
-5 -46 V
-4 -46 V
-5 -47 V
-4 -46 V
-5 -46 V
-4 -46 V
-4 -47 V
-4 -47 V
-4 -46 V
-4 -47 V
-4 -47 V
-3 -47 V
-4 -46 V
0 -2 V
LT2
4553 2609 M
180 0 V
4499 251 M
-42 149 V
-53 183 V
-53 177 V
-54 170 V
-53 164 V
-53 156 V
-53 150 V
-53 143 V
-54 136 V
-53 129 V
-53 122 V
-53 116 V
-53 109 V
-54 102 V
-53 95 V
-53 89 V
-53 81 V
-53 75 V
-54 68 V
-53 61 V
-53 55 V
-53 47 V
-53 41 V
-54 34 V
-53 27 V
-53 21 V
-53 13 V
-53 7 V
-54 0 V
-53 -7 V
-53 -13 V
-53 -21 V
-53 -27 V
-54 -34 V
-53 -41 V
-53 -47 V
-53 -55 V
-53 -61 V
-54 -68 V
-53 -75 V
-53 -81 V
-53 -89 V
-53 -95 V
-54 -102 V
-53 -109 V
-53 -116 V
-53 -122 V
-53 -129 V
-54 -136 V
-53 -143 V
-53 -150 V
-53 -156 V
1691 930 L
1637 760 L
1584 583 L
1531 400 L
1489 251 L
LT3
4553 2509 M
180 0 V
4134 251 M
-5 22 V
-36 154 V
-35 150 V
-35 145 V
-36 141 V
-35 136 V
-36 131 V
-35 127 V
-36 123 V
-35 118 V
-36 113 V
-35 109 V
-36 104 V
-35 100 V
-36 95 V
-35 91 V
-36 86 V
-35 82 V
-35 77 V
-36 72 V
-35 68 V
-36 64 V
-35 59 V
-36 54 V
-35 50 V
-36 45 V
-35 41 V
-36 36 V
-35 32 V
-36 27 V
-35 23 V
-36 18 V
-35 14 V
-35 9 V
-36 4 V
-35 0 V
-36 -4 V
-35 -9 V
-36 -14 V
-35 -18 V
-36 -23 V
-35 -27 V
-36 -32 V
-35 -36 V
-36 -41 V
-35 -45 V
-36 -50 V
-35 -54 V
-35 -59 V
-36 -64 V
-35 -68 V
-36 -72 V
-35 -77 V
-36 -82 V
-35 -86 V
-36 -91 V
-35 -95 V
-36 -100 V
-35 -104 V
-36 -109 V
-35 -113 V
-36 -118 V
-35 -123 V
-35 -127 V
1859 999 L
1824 863 L
1788 722 L
1753 577 L
1717 427 L
1682 273 L
-5 -22 V
LT4
4613 2409 D
3707 2020 D
3324 2686 D
3058 2918 D
2866 2972 D
2728 2931 D
2622 2836 D
2536 2700 D
2473 2537 D
2409 2360 D
2356 2142 D
2313 1924 D
2270 1693 D
2239 1448 D
2207 1190 D
2175 931 D
2153 659 D
LT5
4613 2309 A
3654 2346 A
3420 2659 A
3207 2850 A
3026 2931 A
2856 2972 A
2728 2931 A
2622 2836 A
2526 2686 A
2441 2482 A
2356 2237 A
2292 1965 A
2217 1666 A
2164 1394 A
2111 1163 A
2068 918 A
2026 673 A
1983 441 A
stroke
grestore
end
showpage
}
\put(4493,2309){\makebox(0,0)[r]{res.$\tau^{\normalsize 3}$}}
\put(4493,2409){\makebox(0,0)[r]{res.$\varepsilon^{\normalsize 3}$}}
\put(4493,2509){\makebox(0,0)[r]{$\tau^{\normalsize 2}$}}
\put(4493,2609){\makebox(0,0)[r]{$\varepsilon^{\normalsize 2}$}}
\put(4493,2709){\makebox(0,0)[r]{Pad\'e($\tau$)}}
\put(4493,2809){\makebox(0,0)[r]{Pad\'e($\varepsilon$)}}
\put(4708,351){\makebox(0,0){$\alpha$}}
\put(880,2611){%
\special{ps: gsave currentpoint currentpoint translate
270 rotate neg exch neg exch translate}%
\makebox(0,0)[b]{\shortstack{$f_2^{RW}(\alpha)$}}%
\special{ps: currentpoint grestore moveto}%
}
\put(4856,151){\makebox(0,0){4}}
\put(4324,151){\makebox(0,0){3.5}}
\put(3792,151){\makebox(0,0){3}}
\put(3260,151){\makebox(0,0){2.5}}
\put(2728,151){\makebox(0,0){2}}
\put(2196,151){\makebox(0,0){1.5}}
\put(1664,151){\makebox(0,0){1}}
\put(1132,151){\makebox(0,0){0.5}}
\put(600,151){\makebox(0,0){0}}
\put(540,2972){\makebox(0,0)[r]{2}}
\put(540,2700){\makebox(0,0)[r]{1.8}}
\put(540,2428){\makebox(0,0)[r]{1.6}}
\put(540,2156){\makebox(0,0)[r]{1.4}}
\put(540,1884){\makebox(0,0)[r]{1.2}}
\put(540,1612){\makebox(0,0)[r]{1}}
\put(540,1339){\makebox(0,0)[r]{0.8}}
\put(540,1067){\makebox(0,0)[r]{0.6}}
\put(540,795){\makebox(0,0)[r]{0.4}}
\put(540,523){\makebox(0,0)[r]{0.2}}
\put(540,251){\makebox(0,0)[r]{0}}
\end{picture}}}
\caption{\label{fig5.1}
  MF spectrum for diffusion near an absorbing RW. Comparison of different
 approximation and resummation schemes.}
\end{figure}
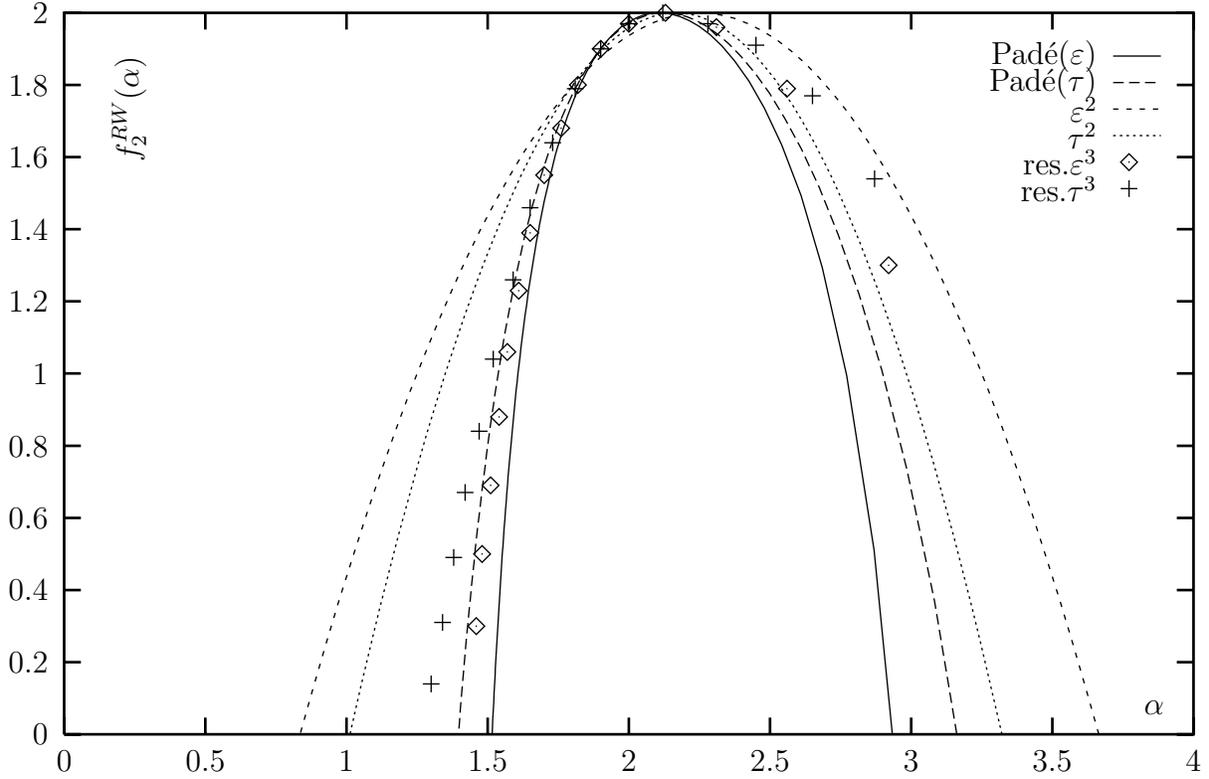

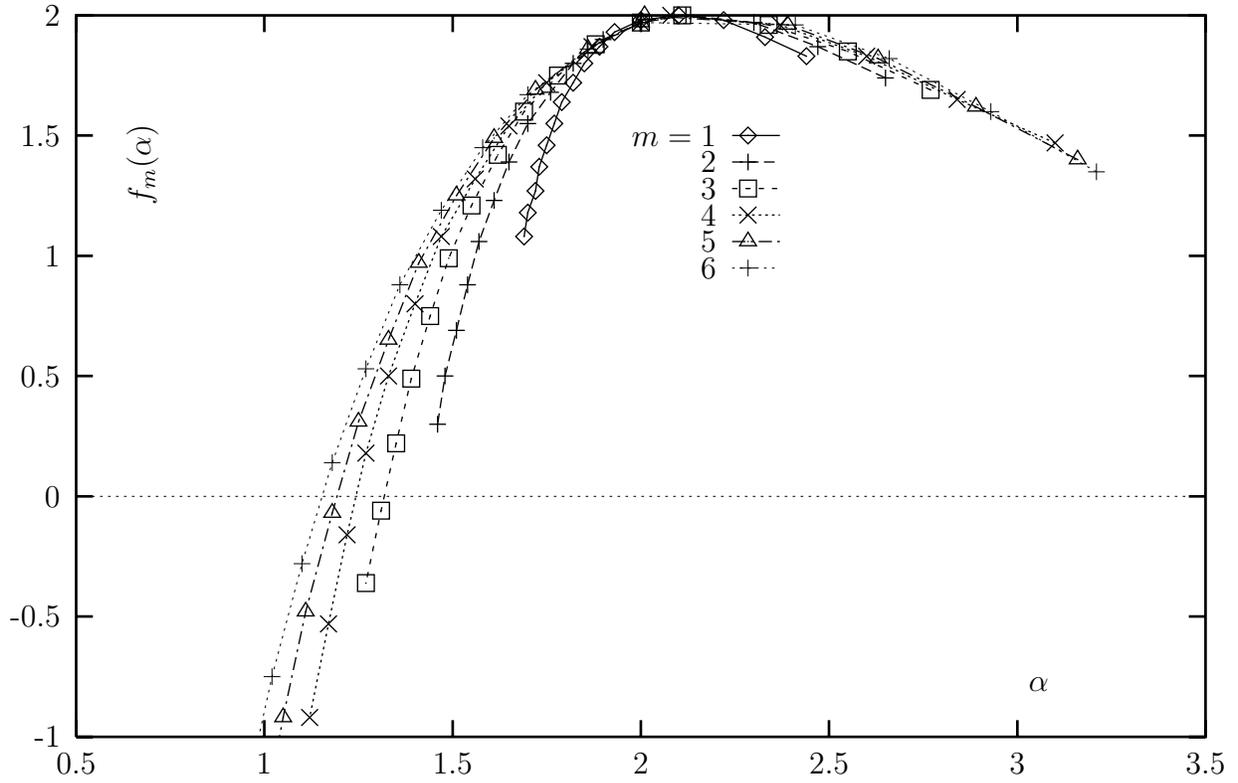
\begin{figure}
\makebox{\hspace*{-20mm}\makebox{
\setlength{\unitlength}{0.1bp}
\special{!
/gnudict 40 dict def
gnudict begin
/Color false def
/Solid false def
/gnulinewidth 5.000 def
/vshift -33 def
/dl {10 mul} def
/hpt 31.5 def
/vpt 31.5 def
/M {moveto} bind def
/L {lineto} bind def
/R {rmoveto} bind def
/V {rlineto} bind def
/vpt2 vpt 2 mul def
/hpt2 hpt 2 mul def
/Lshow { currentpoint stroke M
  0 vshift R show } def
/Rshow { currentpoint stroke M
  dup stringwidth pop neg vshift R show } def
/Cshow { currentpoint stroke M
  dup stringwidth pop -2 div vshift R show } def
/DL { Color {setrgbcolor Solid {pop []} if 0 setdash }
 {pop pop pop Solid {pop []} if 0 setdash} ifelse } def
/BL { stroke gnulinewidth 2 mul setlinewidth } def
/AL { stroke gnulinewidth 2 div setlinewidth } def
/PL { stroke gnulinewidth setlinewidth } def
/LTb { BL [] 0 0 0 DL } def
/LTa { AL [1 dl 2 dl] 0 setdash 0 0 0 setrgbcolor } def
/LT0 { PL [] 0 1 0 DL } def
/LT1 { PL [4 dl 2 dl] 0 0 1 DL } def
/LT2 { PL [2 dl 3 dl] 1 0 0 DL } def
/LT3 { PL [1 dl 1.5 dl] 1 0 1 DL } def
/LT4 { PL [5 dl 2 dl 1 dl 2 dl] 0 1 1 DL } def
/LT5 { PL [4 dl 3 dl 1 dl 3 dl] 1 1 0 DL } def
/LT6 { PL [2 dl 2 dl 2 dl 4 dl] 0 0 0 DL } def
/LT7 { PL [2 dl 2 dl 2 dl 2 dl 2 dl 4 dl] 1 0.3 0 DL } def
/LT8 { PL [2 dl 2 dl 2 dl 2 dl 2 dl 2 dl 2 dl 4 dl] 0.5 0.5 0.5 DL } def
/P { stroke [] 0 setdash
  currentlinewidth 2 div sub M
  0 currentlinewidth V stroke } def
/D { stroke [] 0 setdash 2 copy vpt add M
  hpt neg vpt neg V hpt vpt neg V
  hpt vpt V hpt neg vpt V closepath stroke
  P } def
/A { stroke [] 0 setdash vpt sub M 0 vpt2 V
  currentpoint stroke M
  hpt neg vpt neg R hpt2 0 V stroke
  } def
/B { stroke [] 0 setdash 2 copy exch hpt sub exch vpt add M
  0 vpt2 neg V hpt2 0 V 0 vpt2 V
  hpt2 neg 0 V closepath stroke
  P } def
/C { stroke [] 0 setdash exch hpt sub exch vpt add M
  hpt2 vpt2 neg V currentpoint stroke M
  hpt2 neg 0 R hpt2 vpt2 V stroke } def
/T { stroke [] 0 setdash 2 copy vpt 1.12 mul add M
  hpt neg vpt -1.62 mul V
  hpt 2 mul 0 V
  hpt neg vpt 1.62 mul V closepath stroke
  P  } def
/S { 2 copy A C} def
end
}
\begin{picture}(5039,3023)(0,0)
\special{"
gnudict begin
gsave
50 50 translate
0.100 0.100 scale
0 setgray
/Helvetica findfont 100 scalefont setfont
newpath
-500.000000 -500.000000 translate
LTa
600 1158 M
4256 0 V
LTb
600 251 M
63 0 V
4193 0 R
-63 0 V
600 705 M
63 0 V
4193 0 R
-63 0 V
600 1158 M
63 0 V
4193 0 R
-63 0 V
600 1612 M
63 0 V
4193 0 R
-63 0 V
600 2065 M
63 0 V
4193 0 R
-63 0 V
600 2519 M
63 0 V
4193 0 R
-63 0 V
600 2972 M
63 0 V
4193 0 R
-63 0 V
600 251 M
0 63 V
0 2658 R
0 -63 V
1309 251 M
0 63 V
0 2658 R
0 -63 V
2019 251 M
0 63 V
0 2658 R
0 -63 V
2728 251 M
0 63 V
0 2658 R
0 -63 V
3437 251 M
0 63 V
0 2658 R
0 -63 V
4147 251 M
0 63 V
0 2658 R
0 -63 V
4856 251 M
0 63 V
0 2658 R
0 -63 V
600 251 M
4256 0 V
0 2721 V
-4256 0 V
600 251 L
LT0
3072 2519 M
180 0 V
100 299 R
-156 72 V
-156 64 V
-170 18 V
-142 -18 V
-99 -45 V
-57 -55 V
-57 -63 V
-42 -73 V
-43 -73 V
-28 -81 V
-29 -82 V
-28 -81 V
-14 -91 V
-29 -82 V
-14 -90 V
3132 2519 D
3352 2818 D
3196 2890 D
3040 2954 D
2870 2972 D
2728 2954 D
2629 2909 D
2572 2854 D
2515 2791 D
2473 2718 D
2430 2645 D
2402 2564 D
2373 2482 D
2345 2401 D
2331 2310 D
2302 2228 D
2288 2138 D
LT1
3072 2419 M
180 0 V
398 317 R
-255 118 V
-241 91 V
-256 27 V
-170 -27 V
-142 -64 V
-113 -90 V
-85 -109 V
-86 -118 V
-71 -145 V
-56 -145 V
-57 -155 V
-43 -163 V
-42 -172 V
-43 -172 V
-28 -182 V
3132 2419 A
3650 2736 A
3395 2854 A
3154 2945 A
2898 2972 A
2728 2945 A
2586 2881 A
2473 2791 A
2388 2682 A
2302 2564 A
2231 2419 A
2175 2274 A
2118 2119 A
2075 1956 A
2033 1784 A
1990 1612 A
1962 1430 A
LT2
3072 2319 M
180 0 V
568 372 R
-312 145 V
-298 100 V
-326 36 V
-156 -27 V
-170 -82 V
2416 2745 L
2288 2609 L
-99 -163 V
-99 -191 V
-86 -199 V
-70 -218 V
-71 -236 V
-57 -244 V
-57 -254 V
1692 831 L
3132 2319 B
3820 2691 B
3508 2836 B
3210 2936 B
2884 2972 B
2728 2945 B
2558 2863 B
2416 2745 B
2288 2609 B
2189 2446 B
2090 2255 B
2004 2056 B
1934 1838 B
1863 1602 B
1806 1358 B
1749 1104 B
1692 831 B
LT3
3072 2219 M
180 0 V
1037 272 R
-369 164 V
-341 163 V
-326 118 V
-412 36 V
-113 -27 V
-184 -91 V
2373 2718 L
2231 2555 L
2104 2355 L
1976 2138 L
-99 -254 V
1777 1612 L
-85 -291 V
-71 -308 V
1551 677 L
1480 324 L
3132 2219 C
4289 2491 C
3920 2655 C
3579 2818 C
3253 2936 C
2841 2972 C
2728 2945 C
2544 2854 C
2373 2718 C
2231 2555 C
2104 2355 C
1976 2138 C
1877 1884 C
1777 1612 C
1692 1321 C
1621 1013 C
1551 677 C
1480 324 C
LT4
3072 2119 M
180 0 V
1122 309 R
-383 199 V
-369 182 V
-341 127 V
-539 36 V
-14 -27 V
2529 2845 L
2331 2691 L
2175 2509 L
2033 2292 L
1891 2038 L
1777 1748 L
1664 1439 L
-99 -344 V
1465 723 L
1380 324 L
-14 -73 V
3132 2119 T
4374 2428 T
3991 2627 T
3622 2809 T
3281 2936 T
2742 2972 T
2728 2945 T
2529 2845 T
2331 2691 T
2175 2509 T
2033 2292 T
1891 2038 T
1777 1748 T
1664 1439 T
1565 1095 T
1465 723 T
1380 324 T
LTa
3072 2019 M
180 0 V
1193 363 R
-398 227 V
-383 200 V
-354 127 V
-582 9 V
2529 2845 L
2302 2673 L
2132 2473 L
1976 2237 L
1820 1956 L
1692 1639 L
1565 1285 L
1451 904 L
1338 478 L
1288 251 L
3132 2019 A
4445 2382 A
4047 2609 A
3664 2809 A
3310 2936 A
2728 2945 A
2529 2845 A
2302 2673 A
2132 2473 A
1976 2237 A
1820 1956 A
1692 1639 A
1565 1285 A
1451 904 A
1338 478 A
stroke
grestore
end
showpage
}
\put(3012,2019){\makebox(0,0)[r]{$6$}}
\put(3012,2119){\makebox(0,0)[r]{$5$}}
\put(3012,2219){\makebox(0,0)[r]{$4$}}
\put(3012,2319){\makebox(0,0)[r]{$3$}}
\put(3012,2419){\makebox(0,0)[r]{$2$}}
\put(3012,2519){\makebox(0,0)[r]{$m=1$}}
\put(4228,451){\makebox(0,0){$\alpha$}}
\put(880,2411){%
\special{ps: gsave currentpoint currentpoint translate
270 rotate neg exch neg exch translate}%
\makebox(0,0)[b]{\shortstack{$f_m(\alpha)$}}%
\special{ps: currentpoint grestore moveto}%
}
\put(4856,151){\makebox(0,0){3.5}}
\put(4147,151){\makebox(0,0){3}}
\put(3437,151){\makebox(0,0){2.5}}
\put(2728,151){\makebox(0,0){2}}
\put(2019,151){\makebox(0,0){1.5}}
\put(1309,151){\makebox(0,0){1}}
\put(600,151){\makebox(0,0){0.5}}
\put(540,2972){\makebox(0,0)[r]{2}}
\put(540,2519){\makebox(0,0)[r]{1.5}}
\put(540,2065){\makebox(0,0)[r]{1}}
\put(540,1612){\makebox(0,0)[r]{0.5}}
\put(540,1158){\makebox(0,0)[r]{0}}
\put(540,705){\makebox(0,0)[r]{-0.5}}
\put(540,251){\makebox(0,0)[r]{-1}}
\end{picture}}}
\caption{\label{fig5.2}
   MF spectra for diffusion near an absorbing RW star 
      ($\varepsilon$ expansion).}
\end{figure}

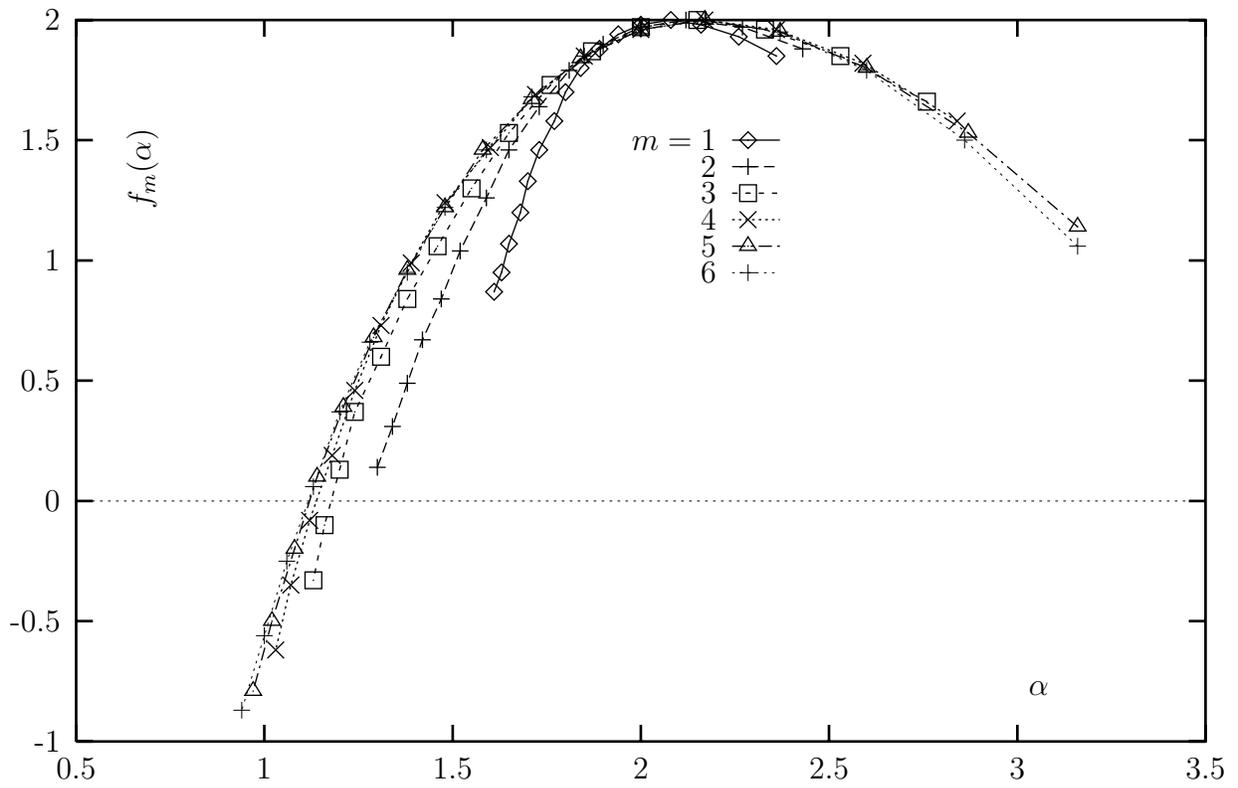
\begin{figure}
\makebox{\hspace*{-20mm}\makebox{
\setlength{\unitlength}{0.1bp}
\special{!
/gnudict 40 dict def
gnudict begin
/Color false def
/Solid false def
/gnulinewidth 5.000 def
/vshift -33 def
/dl {10 mul} def
/hpt 31.5 def
/vpt 31.5 def
/M {moveto} bind def
/L {lineto} bind def
/R {rmoveto} bind def
/V {rlineto} bind def
/vpt2 vpt 2 mul def
/hpt2 hpt 2 mul def
/Lshow { currentpoint stroke M
  0 vshift R show } def
/Rshow { currentpoint stroke M
  dup stringwidth pop neg vshift R show } def
/Cshow { currentpoint stroke M
  dup stringwidth pop -2 div vshift R show } def
/DL { Color {setrgbcolor Solid {pop []} if 0 setdash }
 {pop pop pop Solid {pop []} if 0 setdash} ifelse } def
/BL { stroke gnulinewidth 2 mul setlinewidth } def
/AL { stroke gnulinewidth 2 div setlinewidth } def
/PL { stroke gnulinewidth setlinewidth } def
/LTb { BL [] 0 0 0 DL } def
/LTa { AL [1 dl 2 dl] 0 setdash 0 0 0 setrgbcolor } def
/LT0 { PL [] 0 1 0 DL } def
/LT1 { PL [4 dl 2 dl] 0 0 1 DL } def
/LT2 { PL [2 dl 3 dl] 1 0 0 DL } def
/LT3 { PL [1 dl 1.5 dl] 1 0 1 DL } def
/LT4 { PL [5 dl 2 dl 1 dl 2 dl] 0 1 1 DL } def
/LT5 { PL [4 dl 3 dl 1 dl 3 dl] 1 1 0 DL } def
/LT6 { PL [2 dl 2 dl 2 dl 4 dl] 0 0 0 DL } def
/LT7 { PL [2 dl 2 dl 2 dl 2 dl 2 dl 4 dl] 1 0.3 0 DL } def
/LT8 { PL [2 dl 2 dl 2 dl 2 dl 2 dl 2 dl 2 dl 4 dl] 0.5 0.5 0.5 DL } def
/P { stroke [] 0 setdash
  currentlinewidth 2 div sub M
  0 currentlinewidth V stroke } def
/D { stroke [] 0 setdash 2 copy vpt add M
  hpt neg vpt neg V hpt vpt neg V
  hpt vpt V hpt neg vpt V closepath stroke
  P } def
/A { stroke [] 0 setdash vpt sub M 0 vpt2 V
  currentpoint stroke M
  hpt neg vpt neg R hpt2 0 V stroke
  } def
/B { stroke [] 0 setdash 2 copy exch hpt sub exch vpt add M
  0 vpt2 neg V hpt2 0 V 0 vpt2 V
  hpt2 neg 0 V closepath stroke
  P } def
/C { stroke [] 0 setdash exch hpt sub exch vpt add M
  hpt2 vpt2 neg V currentpoint stroke M
  hpt2 neg 0 R hpt2 vpt2 V stroke } def
/T { stroke [] 0 setdash 2 copy vpt 1.12 mul add M
  hpt neg vpt -1.62 mul V
  hpt 2 mul 0 V
  hpt neg vpt 1.62 mul V closepath stroke
  P  } def
/S { 2 copy A C} def
end
}
\begin{picture}(5039,3023)(0,0)
\special{"
gnudict begin
gsave
50 50 translate
0.100 0.100 scale
0 setgray
/Helvetica findfont 100 scalefont setfont
newpath
-500.000000 -500.000000 translate
LTa
600 1158 M
4256 0 V
LTb
600 251 M
63 0 V
4193 0 R
-63 0 V
600 705 M
63 0 V
4193 0 R
-63 0 V
600 1158 M
63 0 V
4193 0 R
-63 0 V
600 1612 M
63 0 V
4193 0 R
-63 0 V
600 2065 M
63 0 V
4193 0 R
-63 0 V
600 2519 M
63 0 V
4193 0 R
-63 0 V
600 2972 M
63 0 V
4193 0 R
-63 0 V
600 251 M
0 63 V
0 2658 R
0 -63 V
1309 251 M
0 63 V
0 2658 R
0 -63 V
2019 251 M
0 63 V
0 2658 R
0 -63 V
2728 251 M
0 63 V
0 2658 R
0 -63 V
3437 251 M
0 63 V
0 2658 R
0 -63 V
4147 251 M
0 63 V
0 2658 R
0 -63 V
4856 251 M
0 63 V
0 2658 R
0 -63 V
600 251 M
4256 0 V
0 2721 V
-4256 0 V
600 251 L
LT0
3072 2519 M
180 0 V
-13 317 R
-142 73 V
-142 45 V
-114 18 V
-113 -18 V
-85 -36 V
-71 -55 V
-71 -72 V
-57 -91 V
-42 -109 V
-57 -109 V
-43 -118 V
-28 -118 V
-43 -118 V
-28 -108 V
-28 -73 V
3132 2519 D
3239 2836 D
3097 2909 D
2955 2954 D
2841 2972 D
2728 2954 D
2643 2918 D
2572 2863 D
2501 2791 D
2444 2700 D
2402 2591 D
2345 2482 D
2302 2364 D
2274 2246 D
2231 2128 D
2203 2020 D
2175 1947 D
LT1
3072 2419 M
180 0 V
86 444 R
-227 82 V
-213 27 V
-170 -27 V
-142 -64 V
-128 -99 V
2345 2645 L
2231 2482 L
-85 -181 V
-99 -200 V
-71 -181 V
-71 -154 V
-57 -164 V
-56 -163 V
-57 -154 V
3132 2419 A
3338 2863 A
3111 2945 A
2898 2972 A
2728 2945 A
2586 2881 A
2458 2782 A
2345 2645 A
2231 2482 A
2146 2301 A
2047 2101 A
1976 1920 A
1905 1766 A
1848 1602 A
1792 1439 A
1735 1285 A
LT2
3072 2319 M
180 0 V
554 345 R
-326 172 V
-284 100 V
-255 36 V
-213 -27 V
-184 -91 V
2388 2727 L
2231 2546 L
2090 2337 L
1962 2119 L
1848 1920 L
-99 -218 V
-99 -208 V
-57 -218 V
-57 -209 V
1494 859 L
3132 2319 B
3806 2664 B
3480 2836 B
3196 2936 B
2941 2972 B
2728 2945 B
2544 2854 B
2388 2727 B
2231 2546 B
2090 2337 B
1962 2119 B
1848 1920 B
1749 1702 B
1650 1494 B
1593 1276 B
1536 1067 B
1494 859 B
LT3
3072 2219 M
180 0 V
668 372 R
-355 218 V
-326 127 V
-270 36 V
-241 -36 V
2515 2836 L
2331 2691 L
2161 2491 L
1990 2283 L
1863 2056 L
1749 1820 L
-99 -245 V
-85 -245 V
-85 -245 V
1409 841 L
1352 596 L
3132 2219 C
3920 2591 C
3565 2809 C
3239 2936 C
2969 2972 C
2728 2936 C
2515 2836 C
2331 2691 C
2161 2491 C
1990 2283 C
1863 2056 C
1749 1820 C
1650 1575 C
1565 1330 C
1480 1085 C
1409 841 C
1352 596 C
LT4
3072 2119 M
180 0 V
1122 73 R
-412 354 V
-383 245 V
-326 136 V
-284 45 V
-241 -36 V
2501 2827 L
2317 2673 L
2132 2482 L
1990 2265 L
1848 2029 L
1721 1775 L
1607 1512 L
-99 -263 V
1423 977 L
1338 705 L
1267 441 L
3132 2119 T
4374 2192 T
3962 2546 T
3579 2791 T
3253 2927 T
2969 2972 T
2728 2936 T
2501 2827 T
2317 2673 T
2132 2482 T
1990 2265 T
1848 2029 T
1721 1775 T
1607 1512 T
1508 1249 T
1423 977 T
1338 705 T
1267 441 T
LTa
3072 2019 M
180 0 V
1122 100 R
-426 400 V
-369 263 V
-326 145 V
-284 45 V
-241 -36 V
2515 2836 L
2317 2682 L
2146 2482 L
1990 2265 L
1848 2020 L
1707 1757 L
1593 1494 L
-99 -282 V
1394 931 L
1309 650 L
1224 369 L
3132 2019 A
4374 2119 A
3948 2519 A
3579 2782 A
3253 2927 A
2969 2972 A
2728 2936 A
2515 2836 A
2317 2682 A
2146 2482 A
1990 2265 A
1848 2020 A
1707 1757 A
1593 1494 A
1494 1212 A
1394 931 A
1309 650 A
1224 369 A
stroke
grestore
end
showpage
}
\put(3012,2019){\makebox(0,0)[r]{6}}
\put(3012,2119){\makebox(0,0)[r]{5}}
\put(3012,2219){\makebox(0,0)[r]{4}}
\put(3012,2319){\makebox(0,0)[r]{3}}
\put(3012,2419){\makebox(0,0)[r]{2}}
\put(3012,2519){\makebox(0,0)[r]{$m=1$}}
\put(4228,451){\makebox(0,0){$\alpha$}}
\put(880,2411){%
\special{ps: gsave currentpoint currentpoint translate
270 rotate neg exch neg exch translate}%
\makebox(0,0)[b]{\shortstack{$f_m(\alpha)$}}%
\special{ps: currentpoint grestore moveto}%
}
\put(4856,151){\makebox(0,0){3.5}}
\put(4147,151){\makebox(0,0){3}}
\put(3437,151){\makebox(0,0){2.5}}
\put(2728,151){\makebox(0,0){2}}
\put(2019,151){\makebox(0,0){1.5}}
\put(1309,151){\makebox(0,0){1}}
\put(600,151){\makebox(0,0){0.5}}
\put(540,2972){\makebox(0,0)[r]{2}}
\put(540,2519){\makebox(0,0)[r]{1.5}}
\put(540,2065){\makebox(0,0)[r]{1}}
\put(540,1612){\makebox(0,0)[r]{0.5}}
\put(540,1158){\makebox(0,0)[r]{0}}
\put(540,705){\makebox(0,0)[r]{-0.5}}
\put(540,251){\makebox(0,0)[r]{-1}}
\end{picture}}}
\caption{\label{fig5.3}
   MF spectra for diffusion near an absorbing RW star ($\tau$-expansion).}
\end{figure}

The results of the resummation of the series for the spectral functions 
are presented in figures \ref{fig5.1}-\ref{fig5.5}. Each point marked by 
a symbol corresponds to the resummation of both $f_m(\alpha_{mn})$ and 
$\alpha_{mn}$ for a given pair $(m,n)$ where we used a half-integer spacing 
for the values of $n$.
Note that the right wings of the curves correspond to negative $n<0$. In this
region reliable resummations were feasible only for sufficiently large m.
We have only included resummations that were successful in minimizing the
deviation between the second and third order resummed values as described in
the appendix \ref{appB}.

In Fig.\ref{fig5.1} we study the effect of different RG and resummation
procedures for $f^{\rm RW}_2(\alpha)$ in $d=3$ dimensions.
The most simple approach is to directly insert $\varepsilon=1$ into the 
$\varepsilon$-expansion and for the $\tau$-expansion to use 
$\tau=1$ and the $d=3$-dimensional values for the integrals . 
As can be seen from the curves, this will work only for the series
truncated at second order and for small $n$, i.e. near the maximum of 
$f_2(\alpha)$ at $n=0$. In addition we have performed an analytical 
continuation
of our series using [2/1] Pad\'e approximants for the series
truncated at third order. 
The symmetry of the Pad\'e approximant holds only in the region shown and 
may be an artifact of
the method. On the left wing, where it coincides with the resummed
results the Pad\'e approximant gives a continuation which is
compatible with the estimation for the minimal $\alpha$ value
$\alpha_{\min}= d-2$ \cite{Cates}. The Pad\'e result is
$\alpha_{\min}(\varepsilon)=1.333$, $\alpha_{\min}(\tau)=1.017$
for the RW absorber and
$\alpha_{\min}(\varepsilon)=1.250$, $\alpha_{\min}(\tau)=1.013$
for the SAW absorber, which is calculated here only from third order
perturbation theory.
The Pad\'e approximant, while already significantly improving the convergence
of the results, introduces some apparently artificial singularities.
Moreover it does not make full use of the known asymptotics for the
$\varepsilon$-expansions. We have therefore chosen a more sophisticated 
method of 
resummation that has proven to reproduce reliable data in many field theoretic 
applications
\cite{Zinn89,LeGuillou80}. The results of these resummations are again shown 
for both RG approaches.  Note that though the
results obtained for $\alpha_n$ and $f_2(\alpha_n)$ for a specific value of $n$
differ in both approaches, the same curve $f_2(\alpha)$ is described with
better coincidence for the left wing of the curves, corresponding to
positive $n$.

Figs. \ref{fig5.2}-\ref{fig5.5} present the resummed MF spectra 
$f_m(\alpha)$ of Brownian motion near general $m$-leg polymer stars in $d=3$ 
dimensions.
The family of curves $f_m(\alpha)$ appear to approach some limiting 
envelope for increasing $m$ in all cases. This behavior is more pronounced
in the case of Brownian motion near an absorbing SAW star. This provides 
evidence that the MF spectrum catches rather general properties of the 
phenomena 
under consideration. Here, for the absorption of diffusing particles on a 
polymer star the spectrum only slightly varies with the number of legs $m$
of the star even in the vicinity of the core of the star. Only the absorption 
on an endpoint ($m=1$) proves to be an exception.

The behavior of the maximum of the spectra may also be studied in terms of the 
series expansion. The original position of the maximum is given by its 
$\alpha$-coordinate in $\varepsilon$-expansion in the following form:
\begin{eqnarray}
\label{5.1}
\alpha_{m,0} &=&  \eta_{m,1}-\eta'_{m,0},
\\
\label{5.2}
 \alpha_{\rm max}^{\rm RW} &=& m\varepsilon/8 +\ldots,
\\
\label{5.3}
 \alpha_{\rm max}^{\rm SAW} &=& m(1-m)\varepsilon/8 +\ldots.
\end{eqnarray}
For $m>1$ the position of the SAW maximum is shifted in opposite direction to
that of the RW maximum.  
In $\varepsilon$-expansion we find for the curvature at the 
maximum:
\begin{eqnarray}
\label{5.4}
1/f_m''(\alpha) &=& -\eta''_{m,0},
\\
\label{5.5}
 1/f^{\prime\prime\rm RW}_m(\alpha) &=& -m\varepsilon^2/4
     (1-\varepsilon/2(2m-6+3\zeta(3))+\ldots\; ,
\\
\label{5.6}
 1/f^{\prime\prime\rm SAW}_m(\alpha) &=& -9m\varepsilon^2/64 +\ldots\; .
\end{eqnarray}
Here, we use the notations $f_m'(\alpha)=d/d\alpha f_m(\alpha_{m,n})|_{n=0}$ 
and 
$\eta'_{m,n}=d/dn\eta_{m,n}|_{n=0}$ and correspondingly for higher derivatives.
As can be seen also in the plots, the radius ${\cal R}_m\sim 1/f_m''(\alpha)$
of the curvature increases with $m$ for both RW and SAW star. 
Some asymmetry is also present in the plots. It may be more explicitly 
extracted from the series by considering
\begin{eqnarray}
\label{5.7}
 f_m''(\alpha)/f_m'''(\alpha) &=& (\eta_{m,0}'')^{2}/\eta_{0,m}'''\; ,
\\
\label{5.8}
 f^{\prime\prime\rm RW}_m(\alpha)/f^{\prime\prime\prime\rm RW}_m(\alpha) &=& 
    m\varepsilon/12(1-\varepsilon(2m-6+3\zeta(3))+ \ldots \; ,
\\
\label{5.9}
 f^{\prime\prime\rm SAW}_m(\alpha)/f^{\prime\prime\prime\rm SAW}_m(\alpha) &=& 
    m\varepsilon/16(1-\varepsilon(7m/2-175/24+15/64\zeta(3))+ \ldots\; .
\end{eqnarray}
This shows that the asymmetry at the maximum decreases slightly with $m$. The
plots seem to indicate that it approaches some limiting value.

From the plots we present here, in general one may deduce that the series for 
the MF spectra for diffusion near an absorbing polymer star
possess stable resummations and that the shape of the resulting
curves is robust against the change of the number of legs $m$ of the 
polymer star while a limiting curve seems to be approached with increasing $m$.

\section{Conclusions
}\label{VI}

The present work represents extended results on the MF behavior of Brownian 
motion in the vicinity of an absorbing polymer structure. We extend the 
ideas by Cates and Witten \cite{Cates} to map this problem to a problem 
of interacting walks. The former authors used a Fixman expansion technique 
to extract the exponents governing the MF scaling. This approach is equivalent 
to a direct renormalization method and unique to dimensional renormalization 
with $\varepsilon$-expansion. The Fixman expansion assumes without proof the 
renormalizability of the quantities corresponding to higher moments of the 
Laplacian field of Brownian motion. Here, we map the problem to an 
$O(M)$-symmetric 
$\phi^4$-field theory relating the above quantities to corresponding composite 
operators for which renormalizability has been proven \cite{Schaefer92}.
Furthermore, the scaling exponents of these operators have been calculated in 
our previous work \cite{FerHol97b,FerHol97a}. There, it has been shown that 
the resulting spectra of scaling dimensions of these operators display the 
convexity properties which are necessarily found for multifractal scaling but
unusual for power of field operators in field theory 
\cite{Duplantier91,FerHol97}.

The extensive RG study in \cite{FerHol97b,FerHol97a} with three loop results 
allows us here to consider the general case of Brownian motion near the core 
of a star polymer with $m$ legs. The higher order calculation was improved by 
resummation to give reliable estimates for the families of MF spectra 
describing 
the multiscaling of Brownian motion near absorbing RW or SAW stars. 

Our results have proven to be equally stable under change of the general RG 
treatment. We applied two complementary RG approaches, the dimensional 
renormalization with  successive $\varepsilon$-expansion as well as massive 
renormalization at fixed space dimensionalities. The resummation in 
particular  enabled us to extend the region over which the curves for the MF 
spectra coincide in both RG approaches reflecting the stability of the scheme 
of calculations. Our plotted results (figs. \ref{fig5.2}--\ref{fig5.5}) 
indicate some independence of the spectral function on the number of 
legs of the 
absorbing polymer star. For higher number of legs the spectrum seems to 
approach a limiting curve. The MF spectra calculated here show most of 
the common features shared by spectral 
functions that describe a variety of MF phenomena. Let us note however that 
unlike the common definition of the underlying scaling exponents based on 
site averages we here rely on ensemble averages for the moments of the 
Laplacian field of Brownian motion. We average over the configurational 
ensembles of absorbing polymer stars. Only for the case of $m=2$ legs of the 
absorbing star, also a site average definition could be used. As has been 
noted also in \cite{Cates,Halsey96} the ensemble average leads to the 
possibility of negative values of 
the spectral function (see figs. \ref{fig5.1}--\ref{fig5.5}). Furthermore 
the fractal dimension is not defined for the core of a polymer star.

Experimentally such absorbing systems are realized in diffusion controlled
reactions with traps or reaction sites attached to 
polymer chains. This is described  by the irreversible reaction $A+B\to 0$
with freely diffusing particles $A$ and traps $B$ attached along the 
polymer chain \cite{Oshanin}. The higher $n$th moments of the field at some 
special trap might then describe the reaction rate for $A^n + C\to 0$.
If $C$ is located at the core of a polymer star this system realizes 
all aspects of our study.
While extensive MC studies exist for many problems in the field of DLA we 
hope that our detailed calculations might initiate also an MC approach to the 
present problem. The current study also allows an extrapolation to $d=2$ 
dimensions. Though one should not expect to find results of a pure 
two-dimensional approach due to topological restrictions in $d=2$ that are not 
present in the perturbative $\varepsilon$-expansion \cite{note2}.

While standard in field theoretical
studies of critical phenomena, the resummation technique as to our
knowledge was not applied in the theory of multifractals. We hope that
our attempt will attract attention for this possibility in the context
of other problems that arise in the theory of multifractal measures.

\acknowledgements

This work was supported in part by SFB 237 of Deutsche Forschungsgemeinschaft
and by the Minerva foundation. Yu.H. acknowledges hospitality of the Institut 
f\"ur Theoretische Physik II, Heinrich-Heine-Universit\"at D\"usseldorf, where 
this work was completed.

It is our pleasure to acknowledge discussions with Lothar
Sch\"afer, Bertrand Duplantier and Alexander Olemskoi. We are indebted
to Gleb Oshanin for attracting our attention to the references
\cite{Oshanin}.
\begin{appendix}

\section{Three-loop Contributions to the Multifractal Spectrum
}\label{appA}
Here, we collect the three-loop contributions of the expressions for the
H\"older exponents $\alpha_{mn}$
(\ref{4.6}), (\ref{4.8}) and the 
spectral functions $f_m(\alpha_{mn})$ (\ref{4.7}), (\ref{4.9})
obtained in the pseudo-$\varepsilon$-expansion scheme. The expressions read:
\begin{eqnarray}
\label{A.1}
\alpha^{RW}_{3loop}&=&
-\frac{m\varepsilon}{8}
\Big (3i_{{6}}m-3{n}^{2}+3m-6mn+20i_{{1}}
-6i_{{4}}+3i_{{5}}
-3i_{{6}}
+3i_{{7}}
-16{i_{{1}}}^{2}
-46ni_{{1}}
\\&&\nonumber
-5+6i_{{4}}m+3i_{{5}}m+12n-9i_{{4}}{n}^{2}
-12i_{{4}}mn+9{n}^{2}i_{{1}}+32n{i_{{1}}}^{2}-6i_{{7}}n
-6i_{{5}}mn
\\&&\nonumber
-6i_{{6}}mn
+6i_{{6}}n
+18i_{{4}}n-12mi_{{1}}
+24nmi_{{1}}-6i_{{5}}n\Big ),
\\ \label{A.2}
f_{3loop}^{RW}&=&
-\frac{\varepsilon m{n}^{2}}{8}\Big (6-2n-6i_{{4}}n+9i_{{4}}-3i_{{5}}m
-6i_{{4}}m-3i_{{6}}m
-3i_{{7}}-3i_{{5}}+12mi_{{1}}
\\&&\nonumber
+16{i_{{1}}}^{2}
-23i_{{1}}+6ni_{{1}}-3m+3i_{{6}}\Big ),
\\ \label{A.3}
\alpha_{3loop}^{SAW}&=&
-\frac{3m\varepsilon}{512}\Big (-86-718ni_{{1}}-198mi_{{1}}+36i_{{6}}m
+126i_{{4}}m+36i_{{5}}m-27{n}^{2}+54m
\\&&\nonumber
-108mn+332i_{{1}}
+4i_{{2}}-108i_{{4}}+36i_{{5}}-36i_{{6}}+45i_{{7}}-72i_{{5}}n
+72i_{{6}}n-90i_{{7}}n
\\&&\nonumber
+496n{i_{{1}}}^{2}+81{n}^{2}i_{{1}}
-72i_{{5}}mn-81i_{{4}}{n}^{2}-252i_{{4}}mn-72i_{{6}}mn
-8i_{{2}}i_{{1}}+16ni_{{2}}i_{{1}}
\\&&\nonumber
-8ni_{{2}}+396nmi_{{1}}
-248{i_{{1}}}^{2}+190n+270i_{{4}}n\Big ),
\\ \label{A.4}
f_{3loop}^{SAW}&=&
-\frac{3\varepsilon m{n}^{2}}{512}\Big (95-18n-54i_{{4}}n+135i_{{4}}
-36i_{{5}}m-126i_{{4}}m-36i_{{6}}m-45i_{{7}}-36i_{{5}}
\\&&\nonumber
+198mi_{{1}}+248{i_{{1}}}^{2}-359i_{{1}}+54ni_{{1}}-54m
+36i_{{6}}-4i_{{2}}+8i_{{2}}i_{{1}}\Big ).
\end{eqnarray}
The numerical values of the two-loop ($i_1$, $i_2$) and three-loop 
$i_3$--$i_8$ integrals at $d=3$ equal \cite{integrals}: 
$i_1=2/3$, $i_2=-2/27$, 
$i_3=-0.0376820725$,  $i_4=0.3835760966$, $i_5=0.5194312413$,
$i_6=1/2$,  $i_7=0.1739006107$,   $i_8=-0.0946514319$. 
\section{Resummation Procedure
}\label{appB}
Here, we briefly describe the resummation method for the asymptotic 
series that we applied in our calculations 
\cite{LeGuillou80,ass,ZinnJustin81}.  
The starting point is the expansion for the 
function of interest:
\begin{equation}\label{B.1}
\beta(\varepsilon) = \sum_k A_k \varepsilon^k. 
\end{equation}
The coefficients $A_k$ are supposed to posses the following behavior
\begin{equation} \label{B.2}
A_k = c k^{b_0} (-a)^k k! [1 + O(1/k)], \hspace{1em}
k\to\infty
\end{equation}
with known values of constants $c$, $b_0$, $a$. The property (\ref{B.2}) 
indicates the Borel sumability of the series (\ref{B.1}). The Borel 
resummation 
procedure takes into account the asymptotic behavior of the coefficients and
maps the asymptotic series to a convergent one with the same asymptotic limit.
The
procedure is as follows. For the series (\ref{B.1}) we define a Borel-Leroy
transform $f^B(\varepsilon)$ by  
\begin{equation} \label{B.3}
f^B(\varepsilon)= \sum_{j}
\frac{f^{(j)}\varepsilon^j}{\Gamma(j+b+1)},
\end{equation}
with the Euler Gamma function $\Gamma(x)$ and a fit parameter $b$.
Then the initial series may be regained from
\begin{equation} \label{B.4}
f^{\rm res}(\varepsilon)= \int_{0}^{\infty}d t t^b
e^{-t} f^B(\varepsilon t).
\end{equation}
Assuming the behavior of the high order terms (\ref{B.2}) one concludes that
the singularity of the transformed series
$f^B(\varepsilon)$ closest to
the origin is located at the point $(-1/a)$. Conformally mapping the
$\varepsilon$ plane onto a disk of radius 1 while leaving the origin invariant,
$$
w = \frac{(1+a\varepsilon)^{1/2}-1}{(1+a\varepsilon)^{1/2}+1},
\hspace{3em}
\varepsilon= \frac{4}{a} \frac{w}{(1-w)^2},
$$
and substituting this into 
$f^B(\varepsilon)$, and re-expanding in $w$, we receive a series defined on 
the disk with radius 1 in the $w$ plane. This series is then re-substituted 
into Eq.\ (\ref{B.4}). In order to weaken
a possible singularity in the $w$-plane the corresponding expression is
multiplied by  $(1-w)^{\alpha}$ and thus one more parameter $\alpha$
is introduced \cite{ZinnJustin81}.
 In the resummation procedure the value of $a$ is taken
from the known large-order behavior \cite{Schaefer91,ass}
of the $\varepsilon$-expansion series
while $\alpha$ was chosen in our calculations as a
fit parameter defined by the condition of minimal difference between
resummed second order and third order results. The resummation procedure
was seen to be
quite insensitive to the parameter $b$ introduced by the Borel-Leroy
transformation (\ref{B.3}).
The above procedure was applied to both the $\varepsilon$- and 
pseudo-$\varepsilon$-expansion series.
\end{appendix}

\newpage
\section*{Figures}
\vspace{-5mm}
\vspace{-5mm}
\vspace{-5mm}
\begin{figure}
\makebox{\hspace*{-20mm}\makebox{
\setlength{\unitlength}{0.1bp}
\special{!
/gnudict 40 dict def
gnudict begin
/Color false def
/Solid false def
/gnulinewidth 5.000 def
/vshift -33 def
/dl {10 mul} def
/hpt 31.5 def
/vpt 31.5 def
/M {moveto} bind def
/L {lineto} bind def
/R {rmoveto} bind def
/V {rlineto} bind def
/vpt2 vpt 2 mul def
/hpt2 hpt 2 mul def
/Lshow { currentpoint stroke M
  0 vshift R show } def
/Rshow { currentpoint stroke M
  dup stringwidth pop neg vshift R show } def
/Cshow { currentpoint stroke M
  dup stringwidth pop -2 div vshift R show } def
/DL { Color {setrgbcolor Solid {pop []} if 0 setdash }
 {pop pop pop Solid {pop []} if 0 setdash} ifelse } def
/BL { stroke gnulinewidth 2 mul setlinewidth } def
/AL { stroke gnulinewidth 2 div setlinewidth } def
/PL { stroke gnulinewidth setlinewidth } def
/LTb { BL [] 0 0 0 DL } def
/LTa { AL [1 dl 2 dl] 0 setdash 0 0 0 setrgbcolor } def
/LT0 { PL [] 0 1 0 DL } def
/LT1 { PL [4 dl 2 dl] 0 0 1 DL } def
/LT2 { PL [2 dl 3 dl] 1 0 0 DL } def
/LT3 { PL [1 dl 1.5 dl] 1 0 1 DL } def
/LT4 { PL [5 dl 2 dl 1 dl 2 dl] 0 1 1 DL } def
/LT5 { PL [4 dl 3 dl 1 dl 3 dl] 1 1 0 DL } def
/LT6 { PL [2 dl 2 dl 2 dl 4 dl] 0 0 0 DL } def
/LT7 { PL [2 dl 2 dl 2 dl 2 dl 2 dl 4 dl] 1 0.3 0 DL } def
/LT8 { PL [2 dl 2 dl 2 dl 2 dl 2 dl 2 dl 2 dl 4 dl] 0.5 0.5 0.5 DL } def
/P { stroke [] 0 setdash
  currentlinewidth 2 div sub M
  0 currentlinewidth V stroke } def
/D { stroke [] 0 setdash 2 copy vpt add M
  hpt neg vpt neg V hpt vpt neg V
  hpt vpt V hpt neg vpt V closepath stroke
  P } def
/A { stroke [] 0 setdash vpt sub M 0 vpt2 V
  currentpoint stroke M
  hpt neg vpt neg R hpt2 0 V stroke
  } def
/B { stroke [] 0 setdash 2 copy exch hpt sub exch vpt add M
  0 vpt2 neg V hpt2 0 V 0 vpt2 V
  hpt2 neg 0 V closepath stroke
  P } def
/C { stroke [] 0 setdash exch hpt sub exch vpt add M
  hpt2 vpt2 neg V currentpoint stroke M
  hpt2 neg 0 R hpt2 vpt2 V stroke } def
/T { stroke [] 0 setdash 2 copy vpt 1.12 mul add M
  hpt neg vpt -1.62 mul V
  hpt 2 mul 0 V
  hpt neg vpt 1.62 mul V closepath stroke
  P  } def
/S { 2 copy A C} def
end
}
\begin{picture}(5039,3023)(0,0)
\special{"
gnudict begin
gsave
50 50 translate
0.100 0.100 scale
0 setgray
/Helvetica findfont 100 scalefont setfont
newpath
-500.000000 -500.000000 translate
LTa
LTb
600 251 M
63 0 V
4193 0 R
-63 0 V
600 591 M
63 0 V
4193 0 R
-63 0 V
600 931 M
63 0 V
4193 0 R
-63 0 V
600 1271 M
63 0 V
4193 0 R
-63 0 V
600 1611 M
63 0 V
4193 0 R
-63 0 V
600 1952 M
63 0 V
4193 0 R
-63 0 V
600 2292 M
63 0 V
4193 0 R
-63 0 V
600 2632 M
63 0 V
4193 0 R
-63 0 V
600 2972 M
63 0 V
4193 0 R
-63 0 V
1026 251 M
0 63 V
0 2658 R
0 -63 V
1877 251 M
0 63 V
0 2658 R
0 -63 V
2728 251 M
0 63 V
0 2658 R
0 -63 V
3579 251 M
0 63 V
0 2658 R
0 -63 V
4430 251 M
0 63 V
0 2658 R
0 -63 V
600 251 M
4256 0 V
0 2721 V
-4256 0 V
600 251 L
LT0
3214 2462 M
180 0 V
-198 357 R
-213 -17 V
-170 -51 V
-128 -68 V
-127 -85 V
-128 -85 V
-85 -102 V
-85 -102 V
-43 -102 V
-85 -119 V
-42 -102 V
-43 -119 V
-85 -102 V
3274 2462 D
3196 2819 D
2983 2802 D
2813 2751 D
2685 2683 D
2558 2598 D
2430 2513 D
2345 2411 D
2260 2309 D
2217 2207 D
2132 2088 D
2090 1986 D
2047 1867 D
1962 1765 D
LT1
3214 2362 M
180 0 V
483 321 R
-383 102 V
-298 34 V
-255 -17 V
-170 -68 V
-213 -85 V
2388 2530 L
2217 2411 L
2090 2258 L
1962 2105 L
1834 1935 L
1707 1765 L
-86 -171 V
-85 -187 V
1409 1220 L
3274 2362 A
3877 2683 A
3494 2785 A
3196 2819 A
2941 2802 A
2771 2734 A
2558 2649 A
2388 2530 A
2217 2411 A
2090 2258 A
1962 2105 A
1834 1935 A
1707 1765 A
1621 1594 A
1536 1407 A
1409 1220 A
LT2
3214 2262 M
180 0 V
4813 2105 M
-553 357 V
-425 221 V
-341 102 V
-298 34 V
-255 -17 V
-213 -68 V
2515 2632 L
2345 2513 L
2175 2360 L
2004 2190 L
1834 2020 L
1664 1816 L
1536 1594 L
1409 1373 L
1281 1152 L
1153 914 L
3274 2262 B
4813 2105 B
4260 2462 B
3835 2683 B
3494 2785 B
3196 2819 B
2941 2802 B
2728 2734 B
2515 2632 B
2345 2513 B
2175 2360 B
2004 2190 B
1834 2020 B
1664 1816 B
1536 1594 B
1409 1373 B
1281 1152 B
1153 914 B
LT3
3214 2162 M
180 0 V
1334 -6 R
-510 340 V
-426 187 V
-340 102 V
-256 34 V
-213 -17 V
-255 -68 V
2515 2632 L
2345 2496 L
2132 2343 L
1962 2173 L
1792 1969 L
1621 1748 L
1451 1509 L
1281 1254 L
1153 999 L
983 727 L
3274 2162 C
4728 2156 C
4218 2496 C
3792 2683 C
3452 2785 C
3196 2819 C
2983 2802 C
2728 2734 C
2515 2632 C
2345 2496 C
2132 2343 C
1962 2173 C
1792 1969 C
1621 1748 C
1451 1509 C
1281 1254 C
1153 999 C
983 727 C
LT4
3214 2062 M
180 0 V
1207 145 R
-426 306 V
-383 187 V
-340 85 V
-256 34 V
-213 -17 V
-212 -68 V
2558 2632 L
2345 2496 L
2132 2343 L
1962 2156 L
1749 1935 L
1579 1714 L
1409 1458 L
1238 1186 L
1068 914 L
940 608 L
3274 2062 T
4601 2207 T
4175 2513 T
3792 2700 T
3452 2785 T
3196 2819 T
2983 2802 T
2771 2734 T
2558 2632 T
2345 2496 T
2132 2343 T
1962 2156 T
1749 1935 T
1579 1714 T
1409 1458 T
1238 1186 T
1068 914 T
940 608 T
LTa
3214 1962 M
180 0 V
1164 279 R
-426 289 V
-383 170 V
-297 85 V
-256 34 V
-213 -17 V
-212 -68 V
2558 2632 L
2345 2496 L
2132 2343 L
1962 2156 L
1749 1935 L
1579 1697 L
1409 1424 L
1238 1152 L
1068 846 L
898 540 L
3274 1962 A
4558 2241 A
4132 2530 A
3749 2700 A
3452 2785 A
3196 2819 A
2983 2802 A
2771 2734 A
2558 2632 A
2345 2496 A
2132 2343 A
1962 2156 A
1749 1935 A
1579 1697 A
1409 1424 A
1238 1152 A
1068 846 A
898 540 A
stroke
grestore
end
showpage
}
\put(3154,1962){\makebox(0,0)[r]{6}}
\put(3154,2062){\makebox(0,0)[r]{5}}
\put(3154,2162){\makebox(0,0)[r]{4}}
\put(3154,2262){\makebox(0,0)[r]{3}}
\put(3154,2362){\makebox(0,0)[r]{2}}
\put(3154,2462){\makebox(0,0)[r]{$m=1$}}
\put(4228,451){\makebox(0,0){$\alpha$}}
\put(880,2411){%
\special{ps: gsave currentpoint currentpoint translate
270 rotate neg exch neg exch translate}%
\makebox(0,0)[b]{\shortstack{$f_m(\alpha)$}}%
\special{ps: currentpoint grestore moveto}%
}
\put(4430,151){\makebox(0,0){2}}
\put(3579,151){\makebox(0,0){1.8}}
\put(2728,151){\makebox(0,0){1.6}}
\put(1877,151){\makebox(0,0){1.4}}
\put(1026,151){\makebox(0,0){1.2}}
\put(540,2972){\makebox(0,0)[r]{1.8}}
\put(540,2632){\makebox(0,0)[r]{1.6}}
\put(540,2292){\makebox(0,0)[r]{1.4}}
\put(540,1952){\makebox(0,0)[r]{1.2}}
\put(540,1611){\makebox(0,0)[r]{1}}
\put(540,1271){\makebox(0,0)[r]{0.8}}
\put(540,931){\makebox(0,0)[r]{0.6}}
\put(540,591){\makebox(0,0)[r]{0.4}}
\put(540,251){\makebox(0,0)[r]{0.2}}
\end{picture}}}
\caption{\label{fig5.4}
   MF spectra for diffusion near an absorbing SAW star 
      ($\varepsilon$ expansion).}
\end{figure}
\vspace{-5mm}
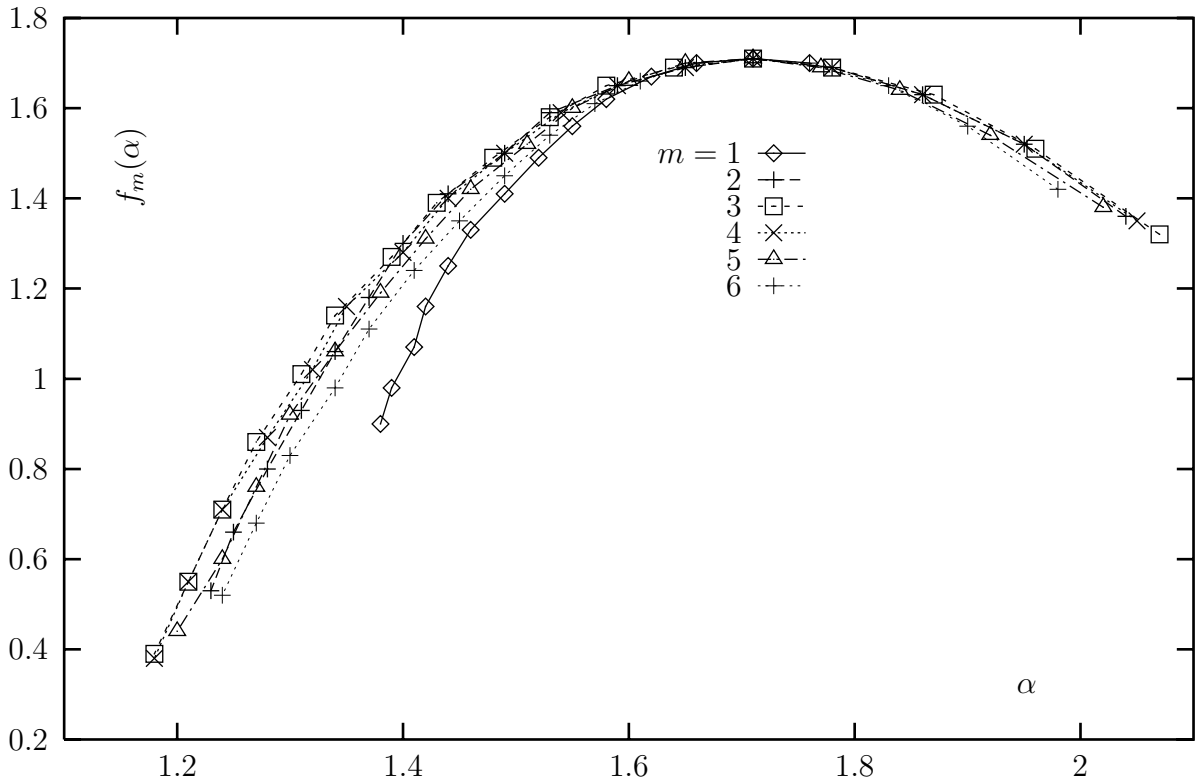
\begin{figure}
\makebox{\hspace*{-20mm}\makebox{
\setlength{\unitlength}{0.1bp}
\special{!
/gnudict 40 dict def
gnudict begin
/Color false def
/Solid false def
/gnulinewidth 5.000 def
/vshift -33 def
/dl {10 mul} def
/hpt 31.5 def
/vpt 31.5 def
/M {moveto} bind def
/L {lineto} bind def
/R {rmoveto} bind def
/V {rlineto} bind def
/vpt2 vpt 2 mul def
/hpt2 hpt 2 mul def
/Lshow { currentpoint stroke M
  0 vshift R show } def
/Rshow { currentpoint stroke M
  dup stringwidth pop neg vshift R show } def
/Cshow { currentpoint stroke M
  dup stringwidth pop -2 div vshift R show } def
/DL { Color {setrgbcolor Solid {pop []} if 0 setdash }
 {pop pop pop Solid {pop []} if 0 setdash} ifelse } def
/BL { stroke gnulinewidth 2 mul setlinewidth } def
/AL { stroke gnulinewidth 2 div setlinewidth } def
/PL { stroke gnulinewidth setlinewidth } def
/LTb { BL [] 0 0 0 DL } def
/LTa { AL [1 dl 2 dl] 0 setdash 0 0 0 setrgbcolor } def
/LT0 { PL [] 0 1 0 DL } def
/LT1 { PL [4 dl 2 dl] 0 0 1 DL } def
/LT2 { PL [2 dl 3 dl] 1 0 0 DL } def
/LT3 { PL [1 dl 1.5 dl] 1 0 1 DL } def
/LT4 { PL [5 dl 2 dl 1 dl 2 dl] 0 1 1 DL } def
/LT5 { PL [4 dl 3 dl 1 dl 3 dl] 1 1 0 DL } def
/LT6 { PL [2 dl 2 dl 2 dl 4 dl] 0 0 0 DL } def
/LT7 { PL [2 dl 2 dl 2 dl 2 dl 2 dl 4 dl] 1 0.3 0 DL } def
/LT8 { PL [2 dl 2 dl 2 dl 2 dl 2 dl 2 dl 2 dl 4 dl] 0.5 0.5 0.5 DL } def
/P { stroke [] 0 setdash
  currentlinewidth 2 div sub M
  0 currentlinewidth V stroke } def
/D { stroke [] 0 setdash 2 copy vpt add M
  hpt neg vpt neg V hpt vpt neg V
  hpt vpt V hpt neg vpt V closepath stroke
  P } def
/A { stroke [] 0 setdash vpt sub M 0 vpt2 V
  currentpoint stroke M
  hpt neg vpt neg R hpt2 0 V stroke
  } def
/B { stroke [] 0 setdash 2 copy exch hpt sub exch vpt add M
  0 vpt2 neg V hpt2 0 V 0 vpt2 V
  hpt2 neg 0 V closepath stroke
  P } def
/C { stroke [] 0 setdash exch hpt sub exch vpt add M
  hpt2 vpt2 neg V currentpoint stroke M
  hpt2 neg 0 R hpt2 vpt2 V stroke } def
/T { stroke [] 0 setdash 2 copy vpt 1.12 mul add M
  hpt neg vpt -1.62 mul V
  hpt 2 mul 0 V
  hpt neg vpt 1.62 mul V closepath stroke
  P  } def
/S { 2 copy A C} def
end
}
\begin{picture}(5039,3023)(0,0)
\special{"
gnudict begin
gsave
50 50 translate
0.100 0.100 scale
0 setgray
/Helvetica findfont 100 scalefont setfont
newpath
-500.000000 -500.000000 translate
LTa
LTb
600 251 M
63 0 V
4193 0 R
-63 0 V
600 591 M
63 0 V
4193 0 R
-63 0 V
600 931 M
63 0 V
4193 0 R
-63 0 V
600 1271 M
63 0 V
4193 0 R
-63 0 V
600 1611 M
63 0 V
4193 0 R
-63 0 V
600 1952 M
63 0 V
4193 0 R
-63 0 V
600 2292 M
63 0 V
4193 0 R
-63 0 V
600 2632 M
63 0 V
4193 0 R
-63 0 V
600 2972 M
63 0 V
4193 0 R
-63 0 V
1026 251 M
0 63 V
0 2658 R
0 -63 V
1877 251 M
0 63 V
0 2658 R
0 -63 V
2728 251 M
0 63 V
0 2658 R
0 -63 V
3579 251 M
0 63 V
0 2658 R
0 -63 V
4430 251 M
0 63 V
0 2658 R
0 -63 V
600 251 M
4256 0 V
0 2721 V
-4256 0 V
600 251 L
LT0
3214 2462 M
180 0 V
15 340 R
-213 17 V
-213 -17 V
-170 -51 V
-170 -85 V
2515 2564 L
2388 2445 L
2260 2309 L
2132 2173 L
-85 -136 V
-85 -153 V
-43 -153 V
-85 -154 V
-42 -136 V
3274 2462 D
3409 2802 D
3196 2819 D
2983 2802 D
2813 2751 D
2643 2666 D
2515 2564 D
2388 2445 D
2260 2309 D
2132 2173 D
2047 2037 D
1962 1884 D
1919 1731 D
1834 1577 D
1792 1441 D
LT1
3214 2362 M
180 0 V
4601 2224 M
-383 272 V
-383 187 V
-341 102 V
-298 34 V
-255 -34 V
-256 -68 V
2430 2615 L
2260 2462 L
2047 2309 L
1877 2122 L
1749 1918 L
1621 1714 L
1494 1492 L
1366 1271 L
1238 1033 L
1153 812 L
3274 2362 A
4601 2224 A
4218 2496 A
3835 2683 A
3494 2785 A
3196 2819 A
2941 2785 A
2685 2717 A
2430 2615 A
2260 2462 A
2047 2309 A
1877 2122 A
1749 1918 A
1621 1714 A
1494 1492 A
1366 1271 A
1238 1033 A
1153 812 A
LT2
3214 2262 M
180 0 V
4728 2156 M
-468 323 V
-383 204 V
-383 102 V
-298 34 V
-298 -34 V
-255 -68 V
2430 2598 L
2217 2445 L
2004 2275 L
1834 2071 L
1621 1850 L
1494 1629 L
1324 1373 L
1196 1118 L
1068 846 L
940 574 L
3274 2262 B
4728 2156 B
4260 2479 B
3877 2683 B
3494 2785 B
3196 2819 B
2898 2785 B
2643 2717 B
2430 2598 B
2217 2445 B
2004 2275 B
1834 2071 B
1621 1850 B
1494 1629 B
1324 1373 B
1196 1118 B
1068 846 B
940 574 B
LT3
3214 2162 M
180 0 V
1249 45 R
-425 289 V
-383 187 V
-341 102 V
-298 34 V
-255 -34 V
-256 -68 V
2473 2615 L
2260 2462 L
2047 2292 L
1877 2088 L
1664 1884 L
1536 1646 L
1366 1390 L
1196 1118 L
1068 846 L
940 557 L
3274 2162 C
4643 2207 C
4218 2496 C
3835 2683 C
3494 2785 C
3196 2819 C
2941 2785 C
2685 2717 C
2473 2615 C
2260 2462 C
2047 2292 C
1877 2088 C
1664 1884 C
1536 1646 C
1366 1390 C
1196 1118 C
1068 846 C
940 557 C
LT4
3214 2062 M
180 0 V
1122 196 R
-426 272 V
-341 170 V
-297 85 V
-256 34 V
-255 -17 V
-213 -68 V
2515 2632 L
2345 2496 L
2132 2326 L
1962 2139 L
1792 1935 L
1621 1714 L
1451 1475 L
1324 1203 L
1196 931 L
1026 659 L
3274 2062 T
4516 2258 T
4090 2530 T
3749 2700 T
3452 2785 T
3196 2819 T
2941 2802 T
2728 2734 T
2515 2632 T
2345 2496 T
2132 2326 T
1962 2139 T
1792 1935 T
1621 1714 T
1451 1475 T
1324 1203 T
1196 931 T
1026 659 T
LTa
3214 1962 M
180 0 V
951 364 R
-340 238 V
-298 153 V
-298 85 V
-213 17 V
-213 -17 V
-212 -68 V
-171 -85 V
2430 2530 L
2260 2377 L
2090 2207 L
1919 2020 L
1749 1799 L
1621 1577 L
1451 1322 L
1324 1067 L
1196 795 L
3274 1962 A
4345 2326 A
4005 2564 A
3707 2717 A
3409 2802 A
3196 2819 A
2983 2802 A
2771 2734 A
2600 2649 A
2430 2530 A
2260 2377 A
2090 2207 A
1919 2020 A
1749 1799 A
1621 1577 A
1451 1322 A
1324 1067 A
1196 795 A
stroke
grestore
end
showpage
}
\put(3154,1962){\makebox(0,0)[r]{6}}
\put(3154,2062){\makebox(0,0)[r]{5}}
\put(3154,2162){\makebox(0,0)[r]{4}}
\put(3154,2262){\makebox(0,0)[r]{3}}
\put(3154,2362){\makebox(0,0)[r]{2}}
\put(3154,2462){\makebox(0,0)[r]{$m=1$}}
\put(4228,451){\makebox(0,0){$\alpha$}}
\put(880,2411){%
\special{ps: gsave currentpoint currentpoint translate
270 rotate neg exch neg exch translate}%
\makebox(0,0)[b]{\shortstack{$f_m(\alpha)$}}%
\special{ps: currentpoint grestore moveto}%
}
\put(4430,151){\makebox(0,0){2}}
\put(3579,151){\makebox(0,0){1.8}}
\put(2728,151){\makebox(0,0){1.6}}
\put(1877,151){\makebox(0,0){1.4}}
\put(1026,151){\makebox(0,0){1.2}}
\put(540,2972){\makebox(0,0)[r]{1.8}}
\put(540,2632){\makebox(0,0)[r]{1.6}}
\put(540,2292){\makebox(0,0)[r]{1.4}}
\put(540,1952){\makebox(0,0)[r]{1.2}}
\put(540,1611){\makebox(0,0)[r]{1}}
\put(540,1271){\makebox(0,0)[r]{0.8}}
\put(540,931){\makebox(0,0)[r]{0.6}}
\put(540,591){\makebox(0,0)[r]{0.4}}
\put(540,251){\makebox(0,0)[r]{0.2}}
\end{picture}}}
\caption{\label{fig5.5}
   MF spectra for diffusion near an absorbing SAW star ($\tau$-expansion).}
\end{figure}

\begin{thebibliography}{99}
\bibitem{note1}
This redefinition leads to some peculiarities in the MF spectrum as discussed
in \cite{Cates} and as we will see later.
\bibitem{note2} E. g. for the 
random walk in $d=2$ the absorbing polymer can in general 
be approached only from one side.
\end{thebibliography}
\end{document}